\documentclass{ectj}

\usepackage{amsfonts,amssymb,graphics,epsfig,verbatim,bm,latexsym,amsmath,url,amsbsy}
\usepackage{bbm,enumerate,xcolor,multirow,graphicx,float}
\usepackage[authoryear]{natbib}

\newtheorem{theorem}{Theorem}
\newtheorem{assumption}{Assumption}
\newtheorem{proposition}{Proposition}

\newtheorem{lemma}{Lemma}
\newtheorem{example}{Example}
\newtheorem{remark}{Remark}

\renewcommand{\thesection}{\arabic{section}}
\renewcommand{\theequation}{\arabic{section}.\arabic{equation}}
\renewcommand{\thetheorem}{\arabic{section}.\arabic{theorem}}
\renewcommand{\theassumption}{\arabic{section}.\arabic{assumption}}
\renewcommand{\theproposition}{\arabic{section}.\arabic{proposition}}

\renewcommand{\thelemma}{\arabic{section}.\arabic{lemma}}

\newcommand{\diag}{{\rm diag}}
\newcommand{\argmin}{\operatorname*{argmin}}

\newcommand{\indf}{\mathbbm{1}}
\newcommand{\rank}{\operatorname*{rank}}
\newcommand{\Xvec}{\boldsymbol{X}}
\newcommand{\xvec}{\boldsymbol{x}}
\newcommand{\Avec}{\boldsymbol{A}}
\newcommand{\Bvec}{\boldsymbol{B}}
\newcommand{\Cvec}{\boldsymbol{C}}
\newcommand{\uvec}{\boldsymbol{u}}
\newcommand{\vvec}{\boldsymbol{v}}
\newcommand{\avec}{\boldsymbol{a}}
\newcommand{\bvec}{\boldsymbol{b}}
\newcommand{\Fvec}{\boldsymbol{f}}
\newcommand{\betavec}{\boldsymbol{\beta}}
\newcommand{\lambdavec}{\boldsymbol{\lambda}}
\newcommand{\alphavec}{\boldsymbol{\alpha}}
\newcommand{\deltavec}{\boldsymbol{\delta}}
\newcommand{\Uvec}{\boldsymbol{U}}
\newcommand{\T}{\mathrm{\scriptscriptstyle{T}}} 
\newcommand{\ci}{\perp\!\!\!\perp}
\newcommand{\Nset}{\mathbb{N}}
\newcommand{\Tset}{\mathbb{T}}
\newcommand{\Yset}{\mathbb{Y}}
\newcommand{\Xset}{\mathbb{X}}
\newcommand{\Aset}{\mathbb{A}}
\newcommand{\Bset}{\mathbb{B}}
\newcommand{\Rset}{\mathbb{R}}
\newcommand{\Dset}{\mathbb{D}}
\newcommand{\Gammavec}{\boldsymbol{\Gamma}}
\newcommand{\gammavec}{\boldsymbol{\gamma}}
\newcommand{\pvec}{\boldsymbol{p}}
\newcommand{\dvec}{\boldsymbol{d}}
\newcommand{\wvec}{\boldsymbol{w}}
\newcommand{\Yvec}{\boldsymbol{Y}}
\newcommand{\Evec}{\boldsymbol{E}}
\newcommand{\Gvec}{\boldsymbol{G}}
\newcommand{\Vvec}{\boldsymbol{V}}
\newcommand{\Mvec}{\boldsymbol{M}}
\newcommand{\Ep}{\mathrm{E}}

\received{...}
\accepted{...}
\volume{1}

\title[Low-Rank Approximations  of Nonseparable Panel Models]{Low-Rank Approximations  of Nonseparable Panel Models}

\author[Fern\'{a}ndez-Val, Freeman, Weidner]{Iv\'{a}n Fern\'{a}ndez-Val$^{\dagger}$,
                Hugo Freeman$^{\ddagger}$ and
                Martin Weidner$^{\S}$}

\address{$^{\dagger}$ Department of Economics, Boston University,
     270 Bay State Road,
     Boston, MA 02215-1403, USA.}
\email{\tt ivanf@bu.edu}

\address{$^{\ddagger}$Department of Economics,
                   University College London,
                   Gower Street,
                   London WC1E~6BT,
                   UK.}
\email{hugo.freeman.16@ucl.ac.uk}

\address{$^{\S}$Nuffield College and Department of Economics, University of Oxford,
   Manor Road, Oxford OX1~3UQ,
                   UK.}
\email{martin.weidner@economics.ox.ac.uk}

\def\AmSTeX{$\cal A$\kern-.1667em\lower.5ex\hbox{$\cal M$}\kern-.125em
    $\cal S$-\TeX}
\def\BibTeX{{\rm B\kern-.05em{\sc i\kern-.025em b}\kern-.08em
    T\kern-.1667em\lower.7ex\hbox{E}\kern-.125emX}}

\begin{document}

    \begin{abstract}
We provide estimation methods for nonseparable panel models based on low-rank factor structure approximations. The factor structures are estimated by matrix-completion methods to deal with the computational challenges of principal component analysis in the presence of missing data. We show that the resulting estimators are consistent in large panels, but suffer from approximation and shrinkage biases. We correct these biases using matching and difference-in-differences approaches. Numerical examples and an empirical application to the effect of election day registration on voter turnout in the U.S. illustrate the properties and usefulness of our methods. 

        \keywords{Nonseparable Panel, Low-Rank Approximations, Matrix Completion, Debias, Two-Way Matching, Election Day Registration}

    \end{abstract}

\section{Introduction}
\setcounter{equation}{0}

Nonseparable models are useful to capture multidimensional unobserved heterogeneity, which is an important feature of economic data. The presence of this heterogeneity makes the effect of covariates on the outcome of interest  different for each unit due to factors that are unobservable or unavailable to the researcher.  In the absence of further restrictions, a different data generating process essentially operates for each unit, which creates identification and estimation challenges.  One way to deal with these challenges is the use of panel data, where each unit is observed on multiple occasions. 
In this paper, we develop an approach to  estimate nonseparable models from panel data based on  homogeneity restrictions and  low-rank factor approximations. Whilst homogeneity restrictions have been used previously in this context, the application of low-rank factor approximations is more novel.

The nonseparable model that we consider includes observed discrete covariates or treatments, multidimensional unobserved individual and time effects, and idiosyncratic errors.  We construct the effects of interest as averages or quantiles of potential outcomes constructed from the model by exogenously manipulating the value of the treatments. These effects are generally not identified from the observed data because the treatment assignment is usually determined by the unobserved  individual and time effects. Following the previous panel literature, we impose cross-section and time-series homogeneity restrictions to identify the effects of interest, see, e.g. \citet{Chamberlain82}, \citet{Manski1987}, \citet{Honore1992}, \citet{evdokimov2010identification}, \citet{GrahamPowell2012}, \citet{HoderleinWhite2012} and \citet{CFHN13}. 

The estimation of the nonseparable model is challenging due to the presence of the multidimensional unobserved individual and time effects. We cannot just exclude these effects because they are endogenous, i.e., related to the treatments. We deal with this problem by approximating their effect with a low-rank factor structure. This approach can be interpreted as a series or sieve approximation on the unobservables.  We characterize the error of this approximation in terms of the functional singular value decomposition of the expectation of the  outcome conditional on the treatment and unobserved effects. For smooth conditional expectation functions, the mean squared error of the approximation error vanishes with the rank of the factor structure at a polynomial rate. 

We develop an estimator of the low-rank factor approximation in the case where the covariate of interest is binary. This  is an empirically relevant case as it covers the treatment effect model for panel data.  We also show how to extend the model to include additive controls and fixed effects. Here, we rely on the  analogy between the estimation of  treatment effects and the matrix completion problem previously noted by \citet{athey17} and \citet{amjad18}.  Thus, given that the principal components program is combinatorially hard in the presence of missing data, we consider the convex relaxation of this program that replaces a constraint in the rank of a matrix by a constraint in its nuclear norm, following \citet{srebro03} and \citet{fazel03}.  The resulting estimator is the matrix-completion estimator. 

The main theoretical result of the paper is to show that the matrix-completion estimator is consistent under asymptotic sequences where the two dimensions of the panel grow to infinity at the same rate. This result does not follow from the existing matrix completion literature that assumes that the matrix to complete has low-rank.  In our case, the underlying matrix of interest can have full rank, but we impose
appropriate smoothness assumptions on the data generating process that guarantee that the  singular values of the matrix
form a rapidly decreasing sequence. This allows a low-rank approximation, and it also implies a bound on the nuclear norm of the 
matrix. Our consistency proof for the matrix completion estimator therefore crucially relies on the bound of the nuclear norm, but does
not impose any low-rank conditions. Our proof strategy also avoids the high-level  \emph{restricted strong convexity} assumption 
  (see e.g.\ \citet{negahban2012restricted}). We instead provide interpretable conditions on the underlying process of the observable and unobservable variables directly.

The matrix-completion estimator is consistent, but can be biased in small samples. This bias comes from two different sources:  approximation bias due to the low-rank factor structure approximation and shrinkage bias due to the nuclear norm regularization of the principal component analysis program \citet{cai10,ma11,bai19joe}. We propose matching approaches to debias the estimator. For each treatment level, the simplest approach consists of finding the observation in  the other treatment level  that is the closest in terms of the estimated factor structure.  We also propose a two-way matching procedure that combines matching with a  differences-in-differences approach. The two-way procedure is related to several recent proposals such as the matching approach of \citet{imai2019use} to estimate causal effects from panel data and the blind regression of \citet{li17b} for matrix completion. The difference with these proposals is in the information used to match the observations.   \citet{imai2019use} use the treatment variable  and   \citet{li17b}  the outcome, whereas we use the estimated factor structure. In this sense, the estimation of the factor structure can be seen as a preliminary de-noising step of the data \citet{chatterjee2015}. \citet{amjad18} proposed a similar debiasing procedure based on the estimated factor structure, but they rely on synthetic control methods instead of matching.  In contemporaneous and independent work, \citet{chlz20} have developed an alternative rotation-debiasing method that can be applied to make inference on heterogenous treatment effects in low-rank models. This method consists of the application of iterative least squares to the left and right singular vectors of the matrix-completion estimator.

We illustrate our methods with an empirical application to the effect of election day registration (EDR) on voter turnout and numerical simulations. We estimate average and quantile effects using a state-level panel dataset on the 24 U.S. presidential elections between 1920 and 2012 collected by \citet{xu17}. We find that, after controlling for possible non-random adoption,  EDR has a positive effect, especially at the bottom of the voter turnout distribution. Our methods uncover stronger effects than standard difference-in-differences methods that rely on restrictive parallel trend assumptions. The simulation results show that our theoretical results provide a good representation of the behavior of the estimators in small samples.

The rest of the paper is organized as follows. Section \ref{sec:model} describes  the model and effects of interest. Section \ref{sec:est} introduces the low-rank factor approximation and derives the properties of its matrix-completion estimator. The matching methods to debias the matrix-completion estimator are discussed in Section \ref{sec:debias}. Section \ref{sec:examples} reports the results of the numerical examples.
All the proofs of the theoretical results are gathered in the Appendix.

\section{Model and Effects of Interest}\label{sec:model}
\setcounter{equation}{0}

Throughout this paper we consider the following nonseparable and nonparametric  panel data model:
\begin{assumption}[Model]\label{ass:model}
\begin{align}
    Y_{it} &= g(\Xvec_{it}, \Avec_i, \Bvec_t, \Uvec_{it} ) ,
    \qquad
    i \in \Nset = \{1,\ldots,N\} ,
    \; 
    t \in \Tset = \{1,\ldots,T\} ,
    \label{model}
\end{align}
where $i$ and $t$ index individual units and time periods, respectively; $Y_{it}$  is an observed outcome or response variable with support  $\Yset \subseteq \mathbb{R}$; $g$ is an unknown function; $\Xvec_{it}$ is a  vector of observed covariates or treatments with  finite support $\Xset$;  $\Avec_i$ 
and $\Bvec_t$ are vectors of individual and time unobserved effects, possibly correlated with $\Xvec_{it}$, with supports $\Aset \subseteq \mathbb{R}^{d_a}$ and $\Bset \subseteq \mathbb{R}^{d_b}$, respectively; and $\Uvec_{it}$ is a vector of unobserved error terms of unspecified dimension,
for which we assume that
\begin{align}
     \Uvec_{it}  \overset{d}{=} \Uvec_{js} \mid \Xvec^{NT}, \Avec^N, \Bvec^T,
     \qquad
     \text{for all } i,j \in \Nset,  \; t,s \in \Tset,
     \label{ass:Main}
\end{align} 
and
\begin{equation}\label{ass:Main2}
 \Uvec_{it} \ci  (\Xvec^{NT}, \Avec^N, \Bvec^T) \mid \Avec_i, \Bvec_t, \qquad
     \text{for all } i  \in \Nset,  \; t  \in \Tset,
\end{equation}
where $\Xvec^{NT} = \{ \Xvec_{it}: i \in \Nset, t \in \Tset\}$, $\Avec^N = \{ \Avec_i: i \in \Nset\}$, $\Bvec^T = \{ \Bvec_t: t \in \Tset\}$, and  $\ci$ denotes stochastic independence.  We also assume that, for all $i  \in \Nset,  \; t  \in \Tset$, the support of $(\Xvec_{it}, \Avec_i, \Bvec_t)$  is 
equal to the Cartesian product $\Xset \times \Aset \times \Bset$, and that $\Ep \, Y_{it}^2 < \infty$.

\end{assumption}

This model can be motivated from a purely statistical perspective as a latent variable model using the Aldous-Hoover  representation for exchangeable random matrices, e.g.  \citet{xu14}, \citet{chatterjee2015}, \citet{or15}, and \citet{li17}.\footnote{In the Aldous-Hoover  representation, $\Avec_i$, $\Bvec_t$ and $\Uvec_{it}$ are independent uniform random variables.} We motivate it instead as a structural model where the unobserved effects $\Avec_i$ and $\Bvec_t$ are associated with individual heterogeneity and aggregate shocks, respectively.  Additional exogenous covariates can be incorporated in the usual way by carrying out the analysis conditional on them. 
We focus on discrete covariates but, from a theoretical perspective, the extension to continuous covariates is 
straightforward by using appropriate smoothing methods --- it is, however, not clear to us whether that extension would be practically useful with realistic sample
sizes. We therefore think that it would complicate our presentation without much benefit.

The main restriction imposed by Assumption \ref{ass:model} is the unit and time homogeneity in \eqref{ass:Main}. A sufficient condition for unit homogeneity is that the observations are identically distributed across $i$, which is a common sampling assumption for panel data.  Time homogeneity has also been commonly used in panel data models \citep{Chamberlain82,Manski1987,Honore1992,evdokimov2010identification,GrahamPowell2012,HoderleinWhite2012,CFHN13}. It implies that time is randomly assigned, conditional on covariates and unobserved effects.  The additional restrictions in \eqref{ass:Main2} are exogeneity conditions on  $(\Xvec^{NT}, \Avec^N, \Bvec^T)$ with respect to  $\Uvec_{it}$, conditional on $\Avec_i$ and $\Bvec_t$. The most substantive is the exogeneity of $\Xvec_{it}$. Given \eqref{ass:Main}, this is a mild condition as time homogeneity already imposes that any relationship between $\Uvec_{it}$ and $\Xvec_{it}$ can only be unit and time-invariant. Taken together, \eqref{ass:Main} and \eqref{ass:Main2} impose that 
\begin{align}\label{ass:Main3}
     \Uvec_{it} \mid \Avec_i, \Bvec_t \overset{d}{=} \Uvec_{js} \mid \Avec_j, \Bvec_s,
     \qquad
     \text{for all } i,j \in \Nset,  \; t,s \in \Tset.     
\end{align}

The product support condition guarantees overlap in the support of the unobserved effects for all values of the treatments. This condition is similar to the overlap condition used  in cross section treatment effect models under unconfoundedness or selection on observables. Thus, together with  \eqref{ass:Main2}, it implies that  $P_{it}(x) :=  \Pr \left( X_{it} = x \mid \Avec^N, \Bvec^T  \right)>0$, a.s., for all $i \in \Nset$, $t \in \Tset$ and $x \in \Xset$, where $P_{it}(x)$ is the analog of the propensity score in our setting. This condition is plausible in many applications. For example, 
  in our empirical application in Section~\ref{sec:Election}, $X_{it}= \indf\{ t \geq \tau_i\} $, where $\tau_i$ 
 is the date of the law change in state $i$. In that case, if we consider $\tau_i$ to be a random variable with sufficiently large support conditional on the unobserved effects, then the condition $ P_{it}(x)>0$, a.s., is satisfied.

The model considered is similar to the static model in \citet{CFHN13}, but there are three important differences.  First, the structural function $g$ has time effects as arguments and therefore allows the relationship between $Y_{it}$ and $\Xvec_{it}$ to vary over time in an unrestricted fashion even under \eqref{ass:Main}. For example, it can include location and scale time effects.  Second, \citet{CFHN13} impose that $Y_{it}$ and $\Xvec_{it}$  are identically distributed across $i$, which is stronger than the unit homogeneity in \eqref{ass:Main3}. Thus, unit homogeneity does not restrict the treatment assignment process.  Third, they analyze short panels, whereas we rely on large $T$ for identification.  Our model also encompasses the nonseparable model with time effects in \citet{freyberger18}, where in our notation $Y_{it} = g_t(\Xvec_{it}, \Avec_i^\T\Bvec_t + \Uvec_{it} )$.\footnote{Note that our model allows for $g$ to depend on $t$ because the dimension of $\Bvec_t$ is unspecified.} 
We provide more examples of models covered by Assumption \ref{ass:model}  below.

The structural function $g$ is generally not identified, but can be used to construct interesting effects. Let $Y_{it}(\xvec) := g(\xvec, \Avec_i, \Bvec_t ,  \Uvec_{it}(\xvec) ) $ be the potential outcome for individual $i$ at time $t$ obtained by setting exogenously $\Xvec_{it} =\xvec \in \Xset$, where
\begin{equation}\label{eq:rs}
\Uvec_{it}(\xvec)  \overset{d}{=} \Uvec_{it} \mid \Avec^N, \Bvec^T.
\end{equation}
Here we impose rank similarity  as the distribution of  $\Uvec_{it}(\xvec)$ conditional on $\Avec^N$ and $\Bvec^T$ does not change with $\xvec$. The main effects of interest are the average structural functions (ASFs) 
\begin{align}\label{eq:asf}
    \mu_{t}(\xvec) := \frac 1 {N} \sum_{i=1}^N  \Ep \left[ Y_{it}(\xvec) \mid \Avec^N, \Bvec^T \right], \ \  \mu(\xvec) :=  \frac 1 {T} \sum_{t=1}^T  \mu_{t}(\xvec),     
\end{align}
and the conditional average structural functions (CASFs)
\begin{align}\label{eq:casf}
    \mu_{t}(\xvec \mid \Xset_0) &:= \frac 1 {N_t(\Xset_0)} \sum_{i=1}^N \indf\{\Xvec_{it} \in \Xset_0 \} \Ep \left[ Y_{it}(\xvec) \mid \Avec^N, \Bvec^T \right], \ \ N_t(\Xset_0) = \sum_{i=1}^N \indf\{\Xvec_{it} \in \Xset_0 \}, \notag \\  
    \mu(\xvec \mid \Xset_0) &:=  \frac 1 {n(\Xset_0)} \sum_{t=1}^T N_t(\Xset_0)  \mu_{t}(\xvec \mid \Xset_0),     \ \ n(\Xset_0) = \sum_{t=1}^T N_t(\Xset_0),
\end{align}
where $\Xset_0 \subseteq \Xset$,  provided that $n(\Xset_0) > 0$.  The ASFs and CASFs correspond to averages of the potential outcome $Y_{it}(\xvec)$ at a given time period or aggregated over the observed time periods. In both cases the average is over the cross sectional units in the observed sample or finite population. Infinite-population versions of the effects can be obtained by taking probability limits as $N \to \infty$. If $\Xvec_{it}$ includes only a binary treatment, the ASFs and CASFs can be used to form treatment effects. For example, $\mu(1) - \mu(0)$ is the time-aggregated average treatment effect and  $\mu_t(1 \,|\, \{1\}) - \mu_t(0 \,|\, \{1\})$ is the average treatment effect on the treated at time $t$.  Distribution structural functions (DSFs) can be constructed analogously  replacing $Y_{it}(\xvec)$ by $\indf \{Y_{it}(\xvec) \leq y\}$ in \eqref{eq:asf} and \eqref{eq:casf} for $y \in \Yset$. Quantile effects can then be formed by taking left-inverses of the DSFs and taking differences. For example, the $\tau$-quantile treatment effect at time $t$ is $q_{t,\tau}(1) - q_{t,\tau}(0)$, where
$$
q_{t,\tau}(\xvec) = \inf\left\{y \in \Yset: \frac 1 {N} \sum_{i=1}^N  \Ep \left[ \indf\{Y_{it}(\xvec) \leq y\} \mid \Avec^N, \Bvec^T \right] \geq \tau \right\}.
$$

We provide some examples of data generating processes that satisfy Assumption \ref{ass:model}. The purpose is to show that Assumption \ref{ass:model} covers a great variety of models commonly used in empirical analysis. Our estimation  methods are generic in that we do not need to specify the data generating process, besides of satisfying Assumption \ref{ass:model}. Of course,  using more information about the data generating process would lead to more efficient estimators, but at the cost of robustness to model misspecification.

\begin{example}[Linear factor model]\label{ex:lfm} \textnormal{Consider the  linear panel model with factor structure in the error terms:
$$
Y_{it}(\xvec) = \xvec^\T \betavec + \lambdavec_i^\T \Fvec_t + \sigma_i(\xvec) \sigma_t(\xvec) U_{it}(\xvec), \ \ U_{it}(\xvec) \mid  \Xvec^{NT}, \Avec^N, \Bvec^T \sim i.i.d. \ F_{U},
$$
where $U_{it}(\xvec)$  is a zero mean random variable with marginal distribution $F_{U}$, which does not depend on $\xvec$.
This is special case of Assumption \ref{ass:model} with $Y_{it} = Y_{it}(\Xvec_{it})$, $\Avec_i = \left(\lambdavec_i, \{\sigma_i(\xvec) : \xvec \in \Xset\}\right),$ $\Bvec_t = \left(\Fvec_t, \{\sigma_t(\xvec) : \xvec \in \Xset\}\right)$, and $\Uvec_{it} = U_{it}(\Xvec_{it})$. The average effect of changing the covariate from $\xvec_0$ to $\xvec_1$ at $t$ is
$$
\mu_t(\xvec_1) - \mu_t(\xvec_0) = \mu_t(\xvec_1 \mid \{\xvec_1\}) - \mu_t(\xvec_0 \mid \{\xvec_1\}) = (\xvec_1 - \xvec_0)^\T \betavec.
$$
A version of this model was considered by \citet{KimOka2014} to analyze the effect of unilateral divorce laws on divorce rates in the U.S.  This model encompasses the standard difference-in-differences model, $Y_{it}(\xvec) = \xvec^\T \betavec + \lambda_i +  f_t + \sigma_i(\xvec) \sigma_t(\xvec) U_{it}(\xvec)$,  by setting $\lambdavec_i = (\lambda_i,1)^{\T}$ and $\Fvec_t = (1,f_t)^{\T}$.}
\end{example}

\begin{example}[Binary response model] \textnormal{Assume that the potential outcome $Y_{it}(\xvec)$ is binary and generated by 
\begin{align*}
      Y_{it}(\xvec) &= \indf\{ m(\xvec,\Avec_i,\Bvec_t) \geq U_{it}(\xvec) \},
      \quad
      U_{it}(\xvec) \mid \Xvec^{NT},\Avec^N,\Bvec^T \sim i.i.d. \, {\cal U}(0,1),
\end{align*}
for some unknown function $m$.   Here, assuming that $U_{it}(\xvec)$ is uniform is a normalization, since $m$ can be arbitrary. This  latent index model with unobserved effects is a special case of Assumption \ref{ass:model} with $Y_{it} = Y_{it}(\Xvec_{it})$ and $\Uvec_{it} = U_{it}(\Xvec_{it})$. The ASFs at  $\xvec$ and $t$ is
$$
\mu_t(\xvec)  =  \frac 1 {N} \sum_{i=1}^N  m(\xvec,\Avec_i,\Bvec_t).
$$
Similar  latent index models for count or censored responses are also covered  by Assumption \ref{ass:model}. }
\end{example}

\begin{example}[Treatment effect factor model]\label{ex:factor} \textnormal{Assume that $\Xvec_{it}$ contains only a binary treatment indicator, i.e., $\Xset = \{0,1\}$.  The potential outcomes are generated by the linear factor model
$$
Y_{it}(\xvec) =  \lambdavec_i(\xvec)^\T \Fvec_t(\xvec) + \sigma_i(\xvec) \sigma_t(\xvec) U_{it}(\xvec), \ \ U_{it}(\xvec) \mid  \Xvec^{NT}, \Avec^N, \Bvec^T \sim i.i.d. \ F_{U}, \ \xvec \in \Xset,
$$
where $U_{it}(\xvec)$  is a zero mean random variable with marginal distribution $F_{U}$, which does not depend on $\xvec$. 
This is special case of Assumption \ref{ass:model} with $
Y_{it} =  Y_{it}(\Xvec_{it})
$, $\Avec_i = \left(\{\lambdavec_i(\xvec), \sigma_i(\xvec) : \xvec \in \Xset\}\right)$, $\Bvec_t = \left(\{\Fvec_t(\xvec), \sigma_t(\xvec) : \xvec \in \Xset\}\right)$, and  $\Uvec_{it} =  U_{it}(\Xvec_{it})$. The average treatment effect  at $t$ is
$$
\mu_t(1) - \mu_t(0)= \frac 1 {N} \sum_{i=1}^N [ \lambdavec_i(1)^\T \Fvec_t(1) - \lambdavec_i(0)^\T \Fvec_t(0)],
$$
and the average effect on the treated at $t$ is
$$
\mu_t(1 \mid \{1\}) - \mu_t(0 \mid \{1\})= \frac 1 {N_t(1)} \sum_{i=1}^N \indf\{\Xvec_{it} = 1 \} [ \lambdavec_i(1)^\T \Fvec_t(1) - \lambdavec_i(0)^\T \Fvec_t(0)],
$$
provided that $N_t(1) = \sum_{i=1}^N \indf\{\Xvec_{it} = 1 \} > 0$. Versions of this model have been  considered by  \citet{hsiao12}, \citet{gobillon16}, \citet{athey17},  \citet{li17}, \citet{xu17}, \citet{li18}, \citet{bai19matrix}, \citet{xiong19}, and \citet{chan20}. 
 Example \ref{ex:lfm} is a special case with $\lambdavec_i(\xvec)^\T \Fvec_t(\xvec) =  \xvec^{\T}\betavec +   \lambdavec_i^\T \Fvec_t$. }
 \end{example}

Throughout this paper we use standard panel data notation, with the two panel dimensions being denoted by units $i$ and time $t$. However,
one could also consider pseudo-panel or network applications of our results, where the two panel dimensions are denoted by $i$ and $j$, and
$Y_{ij}$ could, for example, be wage of worker $i$ in firm $j$, consumption of member $i$ in household $j$, a friendship indicator between individuals $i$ and $j$, or the volume of trade from country $i$ to country $j$.
 The existing literature on two-way heterogeneity in network models usually either makes stronger parametric assumptions than we impose here
(e.g.~\citet{graham2017econometric}, \citet{dzemski2019empirical}, \citet{chen2020nonlinear}, \citet{zeleneev2020identification})
or uses stochastic blockmodels or graphon models, which typically ignore the effect of covariates 
(e.g.~\citet{holland1983stochastic}, \citet{wolfe2013nonparametric}, \citet{gao2015rate}, \citet{auerbach2019identification}). Our methods of estimating non-parametric models
with two-way heterogeneity may therefore also be of interest in a network context.

\section{Estimation via Factor Structure Approximation}\label{sec:est}
\setcounter{equation}{0}

A natural starting point to estimate the effects in \eqref{eq:asf} and \eqref{eq:casf} is to use empirical analogs. This amounts to replacing $ \Ep \left[ Y_{it}(\xvec) \mid \Avec^N, \Bvec^T \right]$ by an estimator. There are two complications with this approach. First, the potential outcome $Y_{it}(\xvec)$ is not observable. We deal with this complication by noting that
\begin{multline*}
\Ep \left[ Y_{it}(\xvec) \mid \Avec^N, \Bvec^T \right] = \Ep \left[ g(\xvec, \Avec_i, \Bvec_t ,  \Uvec_{it}(\xvec) ) \mid \Avec^N, \Bvec^T \right] \\ = \Ep \left[ g(\xvec, \Avec_i, \Bvec_t ,  \Uvec_{it} ) \mid \Avec^N, \Bvec^T \right] = \Ep \left[ g(\xvec, \Avec_i, \Bvec_t ,  \Uvec_{it} ) \mid  \Xvec_{it} = \xvec, \Avec_i, \Bvec_t \right]  \\ = \Ep \left[ Y_{it} \mid \Xvec_{it} = \xvec, \Avec_i, \Bvec_t \right], 
\end{multline*}
 under the rank similarity in \eqref{eq:rs} and  Assumption \ref{ass:model}. Hence, we can write the expectation of the potential outcome as an expectation of the observed outcome. 
The second complication is that $\Avec_i$ and $\Bvec_t$ are not observable, so that we cannot directly estimate  $\Ep \left[ Y_{it} \mid \Xvec_{it} = \xvec, \Avec_i, \Bvec_t \right]$. To deal with this complication, we start by noticing that 
\begin{align}\label{eq:cef}
\begin{split}
    \Ep \left[ Y_{it} \mid \Xvec_{it} = \xvec, \Avec_i = \avec, \Bvec_t  = \bvec \right]  &= \Ep \left[ g(\xvec, \avec, \bvec ,  \Uvec_{it} ) \mid  \Avec_i = \avec, \Bvec_t  = \bvec \right] \\
&=: m(\xvec, \avec, \bvec),
\end{split}
\end{align}
where the function $m$ does not vary with $i$ and $t$, by implication \eqref{ass:Main3} of Assumption \ref{ass:model}. We show next how this function can be approximated and estimated using a low-rank factor structure.

\subsection{Low-rank factor structure approximation}

For ease of exposition, we assume  in the rest of the paper that the covariate vector $\Xvec_{it}$ includes only a  binary treatment and $\Xset = \{0,1\}$. Accordingly, we denote the covariate and its values by $X_{it}$ and $x$ instead of $\Xvec_{it}$ and $\xvec$. In what follows, $x$ denotes a generic element of $\Xset$ and all the assumptions and results hold for all $x \in \Xset_1 \subseteq \Xset$, where  $\Xset_1 = \Xset$ if we are interested in the entire population, $\Xset_1 = \{0\}$ if we are interested in  the treated subpopulation, and $\Xset_1 = \{1\}$ if we are interested in  the untreated subpopulation.

The approximation that we propose is based on the singular value decomposition of the function $(\avec, \bvec) \mapsto m(x,\avec,\bvec) $  for each $x \in \Xset$. We make two assumptions on this decomposition. The first assumption is a sampling condition on the unobserved effects that will be useful to define a norm for the eigenfunctions.

 \begin{assumption}[Sampling of $\Avec_i$ and $\Bvec_t$]~
       \label{ass:SamplingAB}
 (i) $\Avec_i$ is independent and identically distributed across  $i \in \Nset$ with distribution $F_{\Avec}$, (ii) $\Bvec_t$ is independent and identically distributed over $t \in \Tset$  with distribution $F_{\Bvec}$, and (iii) $\Avec_i$ and $\Bvec_t$ are independent for all $i,t$.
\end{assumption}

For simplicity we consider the case where both $\Avec_i$ and $\Bvec_t$ are independently distributed across $i$ and over $t$,
but since we consider asymptotic sequences where both $N$ and $T$ become large one could also allow for
appropriate weak dependence across both $i$ and $t$. Formalizing this weak dependence would complicate both the assumption 
and the proof of the following results, which is why we decided to stick to independence in our presentation here.

The next assumption is a regularity condition on the function $ m(x,\avec,\bvec) $. 
\begin{assumption}[Smoothness of $(\avec,\bvec) \mapsto m(x,\avec, \bvec)$]
     \label{ass:Smoothness}
    The function $(\avec,\bvec) \mapsto m(x,\avec, \bvec)$ admits a singular value decomposition 
     \begin{align*}
          m(x,\avec, \bvec) = \sum_{j=1}^\infty \, s_j(x) \, u_j(x,\avec) \, v_j(x,\bvec) ,
     \end{align*}
     under the $L_2(F_{\Avec} \times F_{\Bvec})$ norm, where the eigenfunctions $u_j(x,\avec)$ and $v_j(x,\bvec)$ are orthonormal, i.e.,
     \begin{align*} 
            \mathbb{E}  \, u_j(x,\Avec_i)^2 &= 1 ,
         &
          \mathbb{E}  \, u_j(x,\Avec_i) u_k(x,\Avec_i)  &=0,\\
            \mathbb{E}  \, v_j(x,\Bvec_t)^2 &= 1,  
          &
            \mathbb{E}  \, v_j(x,\Bvec_t) v_k(x,\Bvec_t)  &=0, 
            & j \neq k&\in \{1,2,3\ldots\},
     \end{align*}
     and the singular values $s_1(x) \geq  s_2(x) \geq  s_3(x) \geq \ldots \geq 0$ satisfy
     \begin{align*}
            \sum_{j=1}^\infty s_j(x) &< \infty.
     \end{align*}
\end{assumption}

There is a large literature on singular value decompositions of functions, which shows that, under appropriate conditions, the singular values satisfy $s_j(x) \lesssim  j^{-\alpha}$,\footnote{ 
Here, $s_j(x) \lesssim  j^{-\alpha}$ means that there exists a constant $c>0$ such that $s_j(x) \leq c\,  j^{-\alpha}$, for all $j$.
} where the decay coefficient $\alpha$
depends on the dimensions of the arguments $\avec$, $\bvec$, and on the smoothness of $(\avec,\bvec) \mapsto m(x,\avec, \bvec) $. For sufficiently smooth functions,
 $\alpha>1$ and therefore $  \sum_{j=1}^\infty s_j(x)  < \infty$. For example, if $(\avec, \bvec) \mapsto m(x,\avec,\bvec)$ is continuously differentiable up to order $s$ and $\Aset$ and $\Bset$ are compact, then 
$$
 s_j(x) \lesssim j^{- \frac{s}{d_a \wedge d_b}  },
$$
by Theorem 3.3 of \citet{gh13}, where $d_a \wedge d_b$ is the minimum of $d_a$ and $d_b$. This implies that $  \sum_{j=1}^\infty s_j(x)  < \infty$ if $s > d_a \wedge d_b $.
Assumption~\ref{ass:Smoothness} is therefore a high-level
smoothness assumption on $(\avec,\bvec) \mapsto m(x,\avec, \bvec)$,
  very similar to the Assumption~2.2. in \citet{menzel2018bootstrap}, where an analogous   condition on the 
singular values is imposed, with the same aim of controlling the behaviour of a function of unobserved two-dimensional heterogeneity.

The formulation of this smoothness assumption is convenient for our purposes, because it immediately leads to a low-rank approximation of  
$m(x,\avec,\bvec)$.
The low-rank  approximation truncates the singular value decomposition to the first $R$ elements,
\begin{equation}\label{eq:approx}
m(x,\avec,\bvec) =  \sum_{j=1}^{\infty} \,  \underbrace{s_j(x)^{1/2} u_j(x,\avec)}_{=:\phi_{j}(x,\avec) }
   \, \underbrace{s_j(x)^{1/2} v_j(x,\bvec) }_{=:  \psi_{j}(x,\bvec)} = \sum_{j=1}^{R} \phi_{j}(x,\avec) \psi_{j}(x,\bvec) + \zeta_R(x,\avec,\bvec).
\end{equation}
The first term is the approximation and the second term is the approximation error. Under Assumption~\ref{ass:Smoothness}, 
$$
\Ep \ \zeta_R(x,\Avec_i,\Bvec_t)^2  \to 0 \ \  \text{ as } \ \ R \to \infty.
$$
In other words, the approximation error can be made negligible by increasing the truncation point $R$. For example, if $s_j(x) \lesssim  j^{-\alpha}$ with $\alpha > 1$, then
\begin{multline*}
 \Ep \ \zeta_R(x,\Avec_i,\Bvec_t)^2 = \Ep\left[ \sum_{j=R+1}^\infty \, s_j(x) \, u_j(x,\Avec_i) \, v_j(x,\Bvec_t) \right]^2 
 \\ = \sum_{j,k =R+1}^\infty s_j(x)s_k(x)  \Ep \left[ u_j(x,\Avec_i)u_k(x,\Avec_i) \right] \, \Ep  \left[ v_j(x,\Bvec_t)v_k(x,\Bvec_t) \right] \\  = \sum_{j =R+1}^\infty s_j(x)^2  \lesssim \sum_{j=R+1}^\infty  j^{-2\alpha}  \leq \int_{R}^{\infty}  j^{-2\alpha} \mathrm{d} j \lesssim  R^{1 - 2 \alpha  },
\end{multline*}
by Assumptions \ref{ass:SamplingAB} and \ref{ass:Smoothness}. 
Hence, $\zeta_R(x,\Avec_i,\Bvec_t)$  converges in mean square to zero at a polynomial rate with $R$.

Combining \eqref{eq:cef} and \eqref{eq:approx}, we obtain the approximate factor model
\begin{equation}\label{eq:afm}
Y_{it} = \lambdavec_{i}(X_{it})^\T \Fvec_{t}(X_{it}) +  \zeta_R(X_{it},\Avec_i,\Bvec_t) + E_{it}, \ \ E_{it} := Y_{it} - \Ep \left[ Y_{it} \mid X_{it}, \Avec_i, \Bvec_t \right],
\end{equation}
where $\lambdavec_{i}(x) = [\phi_{1}(x,\Avec_i), \ldots, \phi_{R}(x,\Avec_i)]^\T$, $\Fvec_{t}(x) = [\psi_{1}(x,\Bvec_t), \ldots, \psi_{R}(x,\Bvec_t)]^\T$, and the composite error $\nu_{it} := \zeta_R(X_{it},\Avec_i,\Bvec_t) + E_{it}$ contains the approximation error, $\zeta_R(X_{it},\Avec_i,\Bvec_t) $, and the conditional expectation error, $E_{it}$. 
 The factor structure  can be seen as a series or sieve approximation to the function $(\avec,\bvec) \mapsto m(x,\avec,\bvec)$  with basis functions $\{\phi_j(x,\avec) \psi_j(x,\bvec)\}_{j=1}^{\infty}$ if we let $R=R_{N,T}$ to grow with $N$ and $T$ such that $\zeta_{R}(x,\avec,\bvec)$ vanishes as $N,T \to \infty$. The factor structure approximation is exact in some cases for fixed $R$. For instance, in Example \ref{ex:factor}
\[
m(x,\Avec_i,\Bvec_t) = \lambdavec_i(x)^\T \Fvec_t(x),
\]
so that $\zeta_R(x,\Avec_i,\Bvec_t)  = 0$, a.s.,  if $R$ is greater or equal to the number of factors.

In the model \eqref{eq:afm} the factor structure changes with the treatment level. In other words, we have a different pure factor model for each $x \in \Xset$, that is
$$
Y_{it} = \lambdavec_{i}(x)^\T \Fvec_{t}(x) + \nu_{it} \text{ if } X_{it} = x.
$$  
This observation leads to our first estimation strategy where the data is partitioned by the treatment level and separate factors and factor loadings are estimated in each element of the partition by solving the least squares program
\begin{equation}\label{eq:pca}
 \min_{\{\lambdavec_{i}\}_{i=1}^N, \{\Fvec_{t}\}_{t=1}^T} \frac{1}{2} \sum_{i=1}^N \sum_{t=1}^T D_{it}(x) \left(Y_{it} - \lambdavec_{i}^\T \Fvec_{t}\right)^2,
\end{equation}
where $D_{it}(x) := \indf\{X_{it} = x\} $. 
Unfortunately, we cannot solve this problem using standard principal component analysis due to the presence of missing data, that is, each observational unit $(i,t)$ is not available at all treatment levels. In the next section, we apply matrix completion methods to deal with this problem.

\subsection{Estimation by matrix completion methods}
We start by expressing the program \eqref{eq:pca} in matrix form. Let $\Gammavec^R(x) = \lambdavec^N(x) \Fvec^T(x)^\T$, where $\lambdavec^N(x) = [\lambdavec_1(x), \ldots, \lambdavec_N(x)]^\T$, a $N \times R$ matrix of factor loadings, and $\Fvec^T(x) = [\Fvec_1(x), \ldots, \Fvec_T(x)]^\T$, a $T \times R$ matrix of factors.  The least squares estimator of $\Gammavec^R(x)$ is the $N \times T$ matrix $\Gammavec $ with typical element $\Gamma_{it}$ that solves
\begin{equation}\label{eq:mc}
 \min_{\{\Gammavec \in \Rset^{N \times T}: \rank(\Gammavec) \leq R \}} \frac{1}{2} \sum_{i=1}^N \sum_{t=1}^T D_{it}(x) \left(Y_{it} - \Gamma_{it} \right)^2.
\end{equation}
Let $\Yvec(x)$ be a $N \times T$ matrix whose $(i,t)$ element is $Y_{it}$ if $X_{it} = x$ and is missing otherwise. The previous program is closely related to the problem of completing the missing entries of $\Yvec(x)$ using a low rank approximation matrix $\Gammavec^R(x)$ \citet{rennie05,candes09,candes10}. This connection was previously noticed by \citet{athey17} and \citet{amjad18} in the context of  treatment effects models. The solution is the $N \times T$ matrix of rank $R$ whose entries are the closest in the mean squared error sense to the corresponding entries of $\Yvec(x)$.

The previous program is combinatorially hard  because of the constraint in the rank of the matrix \citet{srebro03}. 
  Following \citet{fazel03} we consider the convex relaxation of this program.
Let $\|{\bf M}\|_\infty$ be the spectral norm of a $\mathbb{R}^{N \times T}$-matrix ${\bf M}$, and define the nuclear norm (also called trace norm) 
of $\Gammavec$ as the corresponding dual norm
  $\|\Gammavec\|_1 := \max_{\left\{ {\bf M} \in \mathbb{R}^{N \times T} \, : \, \|{\bf M}\|_\infty \leq 1 \right\}} {\rm Tr}\left( {\bf M}' \Gammavec \right)$.
This nuclear norm  can equivalently be defined as the sum of the singular values of $\Gammavec$.
Using this norm we can write the convex relaxation of the program \eqref{eq:mc} as follows,
$$
 \min_{\{\Gammavec \in \Rset^{N \times T}: \|\Gammavec\|_1 \leq R_1 \}} \frac{1}{2}\sum_{i=1}^N \sum_{t=1}^T D_{it}(x) \left(Y_{it} - \Gamma_{it} \right)^2,
$$
where $R_1$ is a positive constant such that $R = f(R_1)$, where $f$ is an increasing function. Hence, $\zeta_R(x,\Avec_i,\Bvec_t)$ vanishes   in mean square   as $R_1 \to \infty$.  We replace the rank constraint, $\rank(\Gammavec) \leq R$, by a constraint on the nuclear norm of the matrix, $\|\Gammavec\|_1 \leq R_1$, i.e. we replace a constraint in the number of nonzero singular values by a constraint in the sum of singular values. This program is convex in $\Gammavec$ and can be reformulated in Lagrange form as
\begin{equation}\label{eq:nn}
 \min_{\{\Gammavec  \in \Rset^{N \times T}\}} \frac{1}{2}\sum_{i=1}^N \sum_{t=1}^T D_{it}(x) \left(Y_{it} - \Gamma_{it} \right)^2 + \rho(R_1) \|\Gammavec\|_1,
\end{equation}
where $\rho(R_1) \geq 0$ is a regularization parameter, which is a one-to-one increasing function of $R_1$. There exist efficient algorithms to solve this program \citet{mazumder10}.

Let $\widehat \Gammavec(x)$ be a solution to \eqref{eq:nn} with typical element $\widehat \Gamma_{it}(x)$. Then,  we can form estimators of the ASF and CASF as
$$
\widehat \mu_{t}(x) = \frac 1 {N} \sum_{i=1}^N  \left[ D_{it}(x) Y_{it} + \{1-D_{it}(x)\} \widehat \Gamma_{it}(x) \right], 
$$
and 
$$
\widehat \mu_{t}(x \mid \{x_0\}) = \frac {\sum_{i=1}^N D_{it}(x_0) \left[ D_{it}(x)Y_{it} + \{1- D_{it}(x) \} \widehat \Gamma_{it}(x)\right]} {\sum_{i=1}^N D_{it}(x_0)}.
$$ 
In the next section, we provide conditions under which these estimators are consistent using asymptotic sequences where $N, T \to \infty$.  These estimators, however, might display shrinkage biases in finite samples due to the nuclear norm regularization \citet{cai10,ma11,bai19joe}. We propose two matching procedures to debias the estimator in Section \ref{sec:debias}.

\subsection{Consistency of Matrix Completion Estimator}\label{sec:mc}

Let $\Gammavec^{\infty}(x)$ be the $N \times T$  matrix with typical element $ \Gamma^{\infty}_{it}(x) = m(x, \Avec_i, \Bvec_t)$ and $\Evec(x)$ be  the $N \times T$ matrix with typical element 
\begin{align}
    E_{it}(x)  &:=  \left\{
    \begin{array}{ll} 
      E_{it} = Y_{it} -   \Gamma^{\infty}_{it}(x)  &  \text{if $X_{it}=x$,}
      \\
        0 & \text{otherwise.}
    \end{array}  
    \right.
    \label{DefE}
\end{align}
Note that $\Gammavec^{\infty}(x) = \lim_{R\to \infty} \Gammavec^{R}(x)$ a.s. 
  Furthermore, we introduce the notation $\mathbb{D}(x) = \{ (i,t) \in \Nset \times \Tset \, : \, X_{it} = x \}$,
and  $n(x)=  |\mathbb{D}(x)|$  for the number of observations with $X_{it} = x$. 

Recall that
\begin{align}
 \widehat \Gammavec(x)
   &\in
 \argmin_{ \Gammavec \in \Rset^{N \times T}}
 Q_{NT}(\Gammavec,\rho,x) ,
 &
 Q_{NT}(  \Gammavec,\rho,x) &= 
  \frac 1 2 \sum_{(i,t) \in \mathbb{D}(x) }
 \left(Y_{it} - \Gamma_{it} \right)^2  + \rho \|  \Gammavec \|_1,
   \label{DefEstimator}
\end{align}
where $\rho := \rho(R_1)$. 
Here, if the $ \argmin$ over $ \Gammavec \in \Rset^{N \times T}  $ is not unique, then we can choose   $ \widehat \Gammavec(x)$
arbitrarily from the set of  minimizers --- our results are not affected by that, we only require that 
$ Q_{NT}( \widehat \Gammavec(x),\rho,x)  \leq  Q_{NT}(\Gammavec,\rho,x) $,
for all $ \Gammavec \in \Rset^{N \times T}$. We want to show that $ \widehat \Gammavec(x)$ converges to $\Gammavec^{\infty}(x)$ as $N,T \to \infty$ in some sense such that $  \widehat  \mu(x)  - \mu(x) = o_P(1)$. For that we require additional assumptions.

\begin{assumption}[Error Moments]~
 \label{ass:SamplingE}
Conditional on $\Xvec^{NT}$, $\Avec^N$ and $\Bvec^T$, $E_{it}(x)$ is independent across $ (i,t) \in \Dset(x) $, 
and there exists a constant  $b < \infty$ that does not depend on $i$, $t$, $N$, $T$, such that
$$
        \Ep\left[E_{it}(x)^4 \mid \Avec^N, \Bvec^T, \Xvec^{NT} \right] \leq b.
$$
Furthermore, we assume that  $n(x)^{-1}   \sum_{(i,t) \in \mathbb{D}(x) } 
     \Gamma^{\infty}_{it}(x)^2 = O_P(1)$.

\end{assumption}

 For the purpose of showing Lemma~\ref{lemma:consistency1} and Theorem~\ref{th:MAIN1} 
we could alternatively replace Assumption~\ref{ass:SamplingE}  by the two high-level conditions:
\begin{align*}
   \frac 2 {n(x)} \sum_{(i,t) \in \mathbb{D}(x) } \Gamma^{\infty}_{it}(x) E_{it}  &= o_P(1) ,
   &
   \| \Evec(x) \|_\infty  &=  O_P\left( \sqrt{N+T} \right),
\end{align*}   
 where again $\|\cdot\|_\infty$ denotes the spectral norm.
 The first of those conditions is implied by Assumption~\ref{ass:SamplingE} 
through application of the weak law of large numbers, 
while the second follows, for example, by the spectral norm inequality in \citet{latala05}.
In principle, we could still derive those high-level conditions if we allowed for appropriate weak dependence of $E_{it}(x)$  across $i$
and over $t$, but we again focus on the independent case  for simplicity of presentation.

We first provide a consistency result for the entries of $\widehat \Gammavec(x)$ that correspond to the observed values of $\Yvec(x)$.  

\begin{lemma}
      \label{lemma:consistency1}
      Let the Assumptions~\ref{ass:SamplingAB}, \ref{ass:Smoothness} and \ref{ass:SamplingE} hold,
      and assume that $\rho = \rho_{NT}$ is chosen such that $\rho_{NT} /  \sqrt{N+T} \rightarrow \infty$ and $\rho_{NT} \sqrt{NT} / n(x) \rightarrow 0$ as $N,T \rightarrow \infty$.
      Then,
     \begin{align*}
            \frac 1 {  n(x)}  \sum_{(i,t) \in \mathbb{D}(x) }
 \left[ \widehat \Gamma_{it}(x)  - \Gamma^{\infty}_{it}(x) \right]^2  
 = o_P(1).
     \end{align*}
\end{lemma}
A necessary condition for the existence of the sequence  $\rho = \rho_{NT}$ in Lemma~\ref{lemma:consistency1}  is 
$n(x) /   \sqrt{(N+T)NT} \rightarrow \infty$, that is, the fraction $n(x) / (NT)$ of  observations with $X_{it} = x$ can converge to zero,
but not too fast. Apart from that, Lemma~\ref{lemma:consistency1} does not restrict the  assignment process that determines $\Xvec^{NT}$.
Notice also that Lemma~\ref{lemma:consistency1} does not require Assumption \ref{ass:model} because $\Gamma^{\infty}(x)$ is a reduced-form parameter. 

Applying  the Cauchy-Schwarz inequality
$$\left( \frac 1 {  n(x)}  \sum_{(i,t) \in \mathbb{D}(x) } a_{it} \right)^2 \leq  \frac 1 {  n(x)}  \sum_{(i,t) \in \mathbb{D}(x) } a_{it}^2$$
with $a_{it} = \widehat \Gamma_{it}(x)  - \Gamma^{\infty}_{it}(x)$, 
Lemma~\ref{lemma:consistency1} 
guarantees that 
$$\frac 1 {  n(x)}  \sum_{(i,t) \in \mathbb{D}(x) }
\left[ \widehat \Gamma_{it}(x)  - \Gamma^{\infty}_{it}(x)  \right]
 = o_P(1).$$
Nevertheless, Lemma~\ref{lemma:consistency1} 
is not directly useful to show the consistency of the estimators of the ASF, because it only guarantees $L_2$-consistency
of $ \widehat \Gammavec(x)$ over the set of entries $(i,t)$ for which $X_{it} = x$. Those are exactly the observations
for which an unbiased estimator of $   \Gamma^{\infty}_{it}(x) =  m(x,\Avec_i, \Bvec_t)$ is already available, namely $Y_{it}$.
The consistency result we would  like to obtain is
\begin{align}
  \frac 1 { NT} \sum_{i=1}^N \sum_{t=1}^T 
 \left[ \widehat \Gamma_{it}(x)  - \Gamma^{\infty}_{it}(x) \right]^2 = o_P(1) ,
    \label{ConsistencyStrong}
 \end{align}
 but such a result will certainly require stronger assumptions on $\Xvec^{NT}$ than we have imposed so far.

 The existing literature on matrix completion relies on the concept
of  \textit{restricted strong convexity} to derive \eqref{ConsistencyStrong}. This approach shows that
under certain conditions on a  $\mathbb{R}^{N \times T}$-matrix $\Mvec$ with entries $M_{it}$, and
on  $\Xvec^{NT}$ (which determines the set $\mathbb{D}(x)$),  there exists a constant $c>0$ such that with high probability
\begin{align*}
      \frac 1 { NT} \sum_{i=1}^N \sum_{t=1}^T  M_{it}^2
    \leq      \frac c {  n(x)}  \sum_{(i,t) \in \mathbb{D}(x) }  M_{it}^2   .
\end{align*}
See Theorem 1 in \citet{negahban2012restricted}, Lemma 12 in \citet{klopp2014noisy}, and Lemma 3 in \citet{athey17}.
Thus, if $M_{it} = \widehat \Gamma_{it}(x)  - \Gamma^{\infty}_{it}(x)$ and $\Xvec^{NT}$  satisfy restricted strong convexity, then \eqref{ConsistencyStrong} would follow from Lemma~\ref{lemma:consistency1}.

We pursue a different strategy than the existing matrix completion literature
to show that
$$
\widehat \mu(x) :=  \frac 1 {T} \sum_{t=1}^T  \widehat \mu_{t}(x)
= \frac 1 {NT} \sum_{i=1}^N \sum_{t=1}^T   D_{it}(x) \, Y_{it} + \frac 1 {NT} \sum_{i=1}^N \sum_{t=1}^T   [1-D_{it}(x) ] \, \widehat \Gamma_{it}(x) 
$$
is a consistent estimator of $(NT)^{-1}  \sum_{i=1}^N \sum_{t=1}^T     \Gamma^{\infty}_{it}$,
which under Assumption~\ref{ass:model} is equal to $\mu(x)$ defined in \eqref{eq:asf}.
We believe that our approach is   simpler in the setting of this paper where
$ \Gamma^{\infty}_{it}(x)$  is not necessarily of low-rank.
In particular, we do not aim to show \eqref{ConsistencyStrong}, but instead we derive consistency of $\widehat \mu(x)$ directly. 
However, the following theorem still requires additional assumptions on the  assignment process that determines $\Xvec^{NT}$, 
in the same way that 
additional conditions on $\Xvec^{NT}$ are required to verify restricted strong convexity.
For simplicity, we focus on consistency of $  \widehat \mu(x)$ in the main text, 
but results for more general weighted averages of the form $ (NT)^{-1}  \sum_{i=1}^N \sum_{t=1}^T   W_{it}(x) \,  \Gamma^{\infty}_{it}(x)$,
with known weights $W_{it}(x) \in \mathbb{R}$,
are presented in the appendix. For example, in the case of the treatment effects on the treated that we consider in the empirical application of Section \ref{sec:Election}, $W_{it}(x) = n(1)^{-1} X_{it}$.
 
\begin{theorem} 
       \label{th:MAIN1}
       Let the Assumptions~\ref{ass:model}, \ref{ass:SamplingAB}, \ref{ass:Smoothness} and \ref{ass:SamplingE} hold.
       Consider $N,T \rightarrow \infty$ at the same rate,
       and let $\rho = \rho_{NT}$ be chosen such that $\rho_{NT} /  \sqrt{N+T} \rightarrow \infty$ 
       and  $\rho_{NT} / \sqrt{NT} \rightarrow 0$.       
       Let
       $  P_{it}(x) =   \Pr \left( X_{it} = x \mid \Avec^N, \Bvec^T  \right)$,
       and assume
       that $ (NT)^{-1} \sum_{i=1}^N \sum_{t=1}^T   P_{it}^{-1}(x)  = O_P(1)$.
       Let $\Gvec(x)$ be the $N \times T$ matrix with entries $ G_{it}(x)  =    P_{it}^{-1}(x) (D_{it}(x) - P_{it}(x)) $,
       and assume that $ \| \Gvec(x) \|_\infty    = O_P(\sqrt{N+T})$, and
       \begin{align}
             \frac 1 {NT}  \sum_{i=1}^N \sum_{t=1}^T  P_{it}^{-1}(x)  \, G_{it}(x) &= o_P(1) ,
             &
               \frac {1} {NT}  \sum_{i=1}^N \sum_{t=1}^T    \Gamma^{\infty}_{it}(x) \, G_{it}(x)  &= o_P(1) .
               \label{ConditionsGit}
       \end{align}
    Then,
    \begin{align*}
         \widehat \mu(x) = \mu(x)  + o_P(1).
    \end{align*}
\end{theorem}

\medskip

To interpret the conditions in Theorem~\ref{th:MAIN1}, notice that
due to the definitions $D_{it}(x) = \indf\{X_{it} = x\} $ and  $  P_{it}(x) =   \Pr \left( X_{it} = x \mid \Avec^N, \Bvec^T  \right)$,  
$\Ep \left[  G_{it}(x)  \mid \Avec^N, \Bvec^T  \right] = 0$ by construction, and $G_{it}(x) $ therefore plays a role very similar
to the error term $E_{it}(x)$. In particular, the conditions in \eqref{ConditionsGit} can  be verified by a weak law of large numbers, as long as $ P_{it}^{-1}(x)$ is not too large, and $G_{it}(x)$ is not too strongly correlated across $i$ and over $t$.
Regarding the condition on the spectral  norm $ \| \Gvec(x) \|_\infty    = O_P(\sqrt{N+T})$, there are many results in the random-matrix
theory literature that show this rate for mean-zero random matrices $\Gvec(x) $, see, for example,  \citet{Geman1980}, \citet{Silverstein1989}, \citet{BaiSilvYin1988},
  \citet{BaiKrishYin1988}. In particular, if $G_{it}(x)$ is independent across both $i$ and $t$, then this rate result follows
 from the very elegant spectral norm inequality in \citet{latala05}, see the proof of Lemma~\ref{lemma:consistency1} in the appendix,
 where we apply that inequality to $E_{it}(x)$. However, that simple argument would require  $X_{it} $ to be independently distributed across $i$ and $t$, 
 conditional on $\Avec^N$, $\Bvec^T$. 
 More generally, we expect $ \| \Gvec(x) \|_\infty    = O_P(\sqrt{N+T})$ to hold whenever
the matrix entries $ G_{it}(x) $ have zero mean, sufficiently bounded moments,
 and weak correlation across both $i$ and $t$, see Section S.2
of the supplementary material of \citet{MoonWeidner2017} for details.

We have thus shown that consistent estimates for ASFs can be obtained via the matrix completion estimator
even if the estimand  $ \Gamma^{\infty}_{it}(x) = m(x, \Avec_i, \Bvec_t)$ itself is not of low rank.
This is the main technical result of this paper. However, inference on $\mu(x)$ based on $\widehat \mu(x) $ can be problematic, because $\widehat \mu(x) $ is subject to both
 low-rank approximation and  shrinkage biases.
The low-rank approximation bias is due to the approximation error $ \zeta_R(x,\avec,\bvec)$
in the decomposition of $m(x,\avec,\bvec)$  in equation \eqref{eq:approx}.
The  shrinkage bias comes from bias in 
 $\widehat \Gammavec(x)$    due to the presence of the nuclear norm penalization in the objective function of \eqref{DefEstimator}.  To isolate this bias, consider a simple case where $Y_{it}(x)$ follows a deterministic pure factor model
$$
Y_{it}(x) = \Gamma_{it}(x) = \sum_{j=1}^R s_j(x) u_j(x,\Avec_i) v_j(x,\Bvec_i).
$$ 
Then, the matrix completion estimator of $ \Gamma_{it}(x)$ in \eqref{DefEstimator} yields
$$
 \widehat \Gamma_{it}(x) =  \sum_{j=1}^R [s_j(x) - \rho]_{+} u_j(x,\Avec_i) v_j(x,\Bvec_i)
$$
where $[z]_{+} = \max(z,0)$.  Compared to $\Gammavec(x)$,  $\widehat \Gammavec(x)$  has the same eigenvectors but the singular values are shrunk toward zero.  This argument carries over to the case where $Y_{it}(x)$ follows an approximate factor structure \citet{cai10,ma11,bai19joe}. Because of these biases, we explore alternative estimates  for $\mu(x)$ in Section~\ref{sec:debias}.

\subsection{Covariates and fixed effects}

As we mentioned in Section \ref{sec:model}, exogenous covariates can be incorporated by conditioning on their values. This method can produce very noisy estimators in small samples unless the covariates take only on few values. Here we consider a semiparametric version of the model that imposes additivity in the effect of the exogenous covariates,  which may be continuous, discrete or mixed. It also allows for additive unobserved individual and time effects that might vary across the covariate level $x$. These effects can be subsumed in the factor structure, but are usually considered separately in empirical analysis as the estimators perform better without regularizing them \citet{athey17}.

Let $\Cvec_{it}$ be a $d_c$-vector of covariates, $\alphavec(x) = (\alpha_1(x),\ldots, \alpha_N(x))$ be a $N$-vector of individual effects and $\deltavec(x) = (\delta_1(x), \ldots, \delta_T(x))$ be a $T$-vector of time effects. Then, we can replace the program  \eqref{eq:nn}  by
\begin{equation*}
  \min_{\{\betavec \in \Rset^{d_c}, \alphavec \in \Rset^N, \deltavec \in \Rset^T, \Gammavec  \in \Rset^{N \times T}\}} \sum_{i=1}^N \sum_{t=1}^T \indf\{X_{it} = x\} \left(Y_{it} - \Cvec_{it}^\T \betavec - \alpha_i - \delta_t -  \Gamma_{it} \right)^2 + \rho(R_1) \|\Gammavec\|_1,
\end{equation*}
\citet{chernozhukov18}, \citet{moon18} and \citet{beyhum19} provide algorithms to solve this program.
Let $\widehat \betavec(x)$,  $\widehat \alphavec(x) = (\widehat \alpha_1(x),\ldots, \widehat \alpha_N(x))$, $\widehat \deltavec(x) = (\widehat \delta_1(x), \ldots, \widehat \delta_T(x))$, and $\widehat \Gammavec(x)$ be the solution of the previous program. We can form estimators of the ASF and CASF as
$$
\widehat \mu_{t}(x) = \frac 1 {N} \sum_{i=1}^N  \left[ \indf\{X_{it} = x\} Y_{it} + \indf\{X_{it} \neq x\} \left\{ \Cvec_{it}^\T \widehat \betavec(x) + \widehat \alpha_i(x) + \widehat \delta_t(x) + \widehat \Gamma_{it}(x) \right\} \right], 
$$
and
\begin{multline*}
\widehat \mu_{t}(x \mid \{x_0\}) = \\  \frac {\sum_{i=1}^N \left[ \indf\{X_{it} = x_0 = x\} Y_{it} + \indf\{X_{it} = x_0 \neq x\} \left\{ \Cvec_{it}^\T \widehat \betavec(x) + \widehat \alpha_i(x) + \widehat \delta_t(x) + \widehat \Gamma_{it}(x) \right\} \right]} {\sum_{i=1}^N \indf\{X_{it} = x_0 \}}.
\end{multline*}

\section{Debiasing Using Matching Methods}\label{sec:debias}
\setcounter{equation}{0}

The matrix completion estimator of the ASF is generally biased. As we explained in Section \ref{sec:mc}, the bias comes from two sources: low-rank approximation  bias and shrinkage bias. One could attempt to correct the shrinkage bias by  shifting the singular values of $\widehat \Gammavec(x)$ upwards. However, inference results on the ASFs based on matrix completion are generally very difficult to obtain even if $ \Gammavec^{\infty}(x)$ is truly low rank. In our setting, the presence of the additional  low-rank approximation bias makes this even more challenging. We instead discuss alternative estimators and show that they have significantly lower biases than the matrix completion estimators in the numerical simulations of Section \ref{sec:MC}. 

To construct the estimators of $ \Gammavec^{\infty}(x)$, we start by extracting the factor structure of $\widehat \Gammavec(x)$   in \eqref{DefEstimator}. Let $ \widehat \lambdavec_i(x)$ and $\widehat  \Fvec_t(x)$ be the $R \times 1$ vectors that satisfy
$$
     \widehat \Gamma_{it}(x) =  \widehat \lambdavec_i(x)^\T  \, \widehat  \Fvec_t(x) ,
$$
subject to the usual normalizations  that $T^{-1} \sum_{t=1}^T \widehat  \Fvec_t(x)  \, \widehat  \Fvec_t(x)^\T  $ is the identity 
matrix of size $R$ and
  $N^{-1} \sum_{i=1}^N  \widehat \lambdavec_i(x) \,  \widehat \lambdavec_i(x)^\T   $ is a diagonal matrix. Next, we apply a matching procedure to this factor structure.  In its simplest version, we estimate each entry $ \Gammavec_{it}^{\infty}(x)$ such that $X_{it} \neq x$, by matching with the observation with $X_{js} = x$ that is the nearest neighbor  in terms of the vectors $\widehat \lambdavec_i(x)$  and $\widehat \Fvec_t(x)$. In particular,  $ \breve \Gamma_{it}(x) =   Y_{i^{**}(i,t,x),t^{**}(i,t,x)}$ where  $i^{**}(i,t,x) \in \Nset$ and $t^{**}(i,t,x) \in \Tset$ are a solution to the program
 \begin{eqnarray*}
   &\min_{j \in \Nset, s \in \Tset}& \left\| \widehat \lambdavec_i(x) - \widehat \lambdavec_j(x) \right\|^2 + \left\| \widehat  \Fvec_t(x) - \widehat \Fvec_s(x) \right\|^2 \\
  &\text{s.t.} &  X_{js} =x.
 \end{eqnarray*}

 We also consider a two-way matching procedure that combines matching with a difference-in-differences  approach. It consists of two steps:
\begin{itemize}
  \item[(i)]   For all $x \in \Xset$ and $(i,t) \in \Nset \times \Tset$ such that $X_{it} \neq x$,  find the matches $i^*(i,t,x) \in \Nset$  and $t^*(i,t,x) \in \Tset$ that solve the program
  \begin{eqnarray*}
  &\min_{j \in \Nset, s \in \Tset}& \left\| \widehat \lambdavec_i(x) - \widehat \lambdavec_j(x) \right\|^2 + \left\| \widehat  \Fvec_t(x) - \widehat \Fvec_s(x) \right\|^2 \\
  &\text{s.t.} & X_{is} = X_{jt} =  X_{js} =x.
  \end{eqnarray*}
          \item[(ii)] Estimate $ \Gamma_{it}(x)$ by
          $$
          \widetilde \Gamma_{it}(x) =  Y_{i,t^*(i,t,x)} + Y_{i^*(i,t,x),t} - Y_{i^*(i,t,x),t^*(i,t,x)}.
          $$
  \end{itemize}
 In other words, we find the match $(j,s)$ with $X_{js} = x$ that not only is the closest to $(i,t)$ in terms of the estimated factor structure, but also corresponds to a unit $j$ with $X_{jt} = x$ and a time period $s$ with $X_{is} = x$.  Then, we estimate the counterfactual $ \Gamma_{it}(x)$ as a linear combination of $Y_{jt}$, $Y_{is}$ and $Y_{js}$. 
 
 The additional  difference-in-differences step in the two-way procedure is useful to reduce bias.   To see this, we can compare $ \widetilde \Gamma_{it}(x)$ with the simple matching estimator $\breve \Gamma_{it}(x)$. 
  Thus, abstracting from the estimation error in the factors and loadings,
\begin{multline*}
\Ep[\breve \Gamma_{it}(x) - \Gamma_{it}(x) \mid \Avec^N, \Bvec^T, \Xvec^{NT}] = m(x,\Avec_{i^{**}(i,t,x)},\Bvec_{t^{**}(i,t,x)}) - m(x,\Avec_i,\Bvec_t) \\
  = \mathcal{O}_P(\|\Avec_{i^{**}(i,t,x)} - \Avec_i\| + \|\Bvec_{t^{**}(i,t,x)} - \Bvec_t\|),
\end{multline*} 
by a first-order Taylor expansion of $(\avec_{i}, \bvec_{t}) \mapsto m(x,\avec_{i},\bvec_{t})$  around $(\Avec_i,\Bvec_t)$; whereas
\begin{multline*}
 \Ep[\widetilde \Gamma_{it}(x) - \Gamma_{it}(x) \mid \Avec^N, \Bvec^T, \Xvec^{NT} ] = m(x,\Avec_{i^{*}(i,t,x)},\Bvec_{t^{*}(i,t,x)}) - m(x,\Avec_i,\Bvec_t)  \\
 = \mathcal{O}_P(\|\Avec_{i^{*}(i,t,x)} - \Avec_i\|^2 + \|\Bvec_{t^{*}(i,t,x)} - \Bvec_t\|^2),
\end{multline*}
by a second-order Taylor expansion of $(\avec_{i}, \bvec_{t}) \mapsto m(x,\avec_{i},\bvec_{t})$  around $(\Avec_i,\Bvec_t)$.  The two-way matching removes the leading term of the Taylor expansion, reducing the bias of the matching by one order of magnitude because $i^{**}(i,t,x) \neq i$ or $t^{**}(i,t,x) \neq t$.  On the other hand, $\|\Avec_{i^{*}(i,t,x)} - \Avec_i\| \geq \|\Avec_{i^{**}(i,t,x)} - \Avec_i\|$ and $\|\Bvec_{t^{*}(i,t,x)} - \Bvec_t\| \geq \|\Bvec_{t^{**}(i,t,x)} - \Bvec_t\|$ a.s. because the two-way procedure imposes the additional restrictions $X_{is} = X_{jt} =x$. Whether the first or second order bias dominates  would generally be determined by the proportion of observations with $X_{js} =x$ and the distributions of $\Avec_i$ and $\Bvec_t$.  We provide a numerical comparison of the biases of the matching estimators in Section~\ref{sec:MC}.

We develop the theory for a  debiased estimator that allows for multiple matches and estimated factors and loadings.  Multiple matches are expected to reduce dispersion at the cost of increasing bias. 
Let $\lambdavec_i = \lambdavec(x,\Avec_i)$ and  $\Fvec_t = \Fvec(x,\Bvec_t)$ be the transformations of $\Avec_i$ 
and $\Bvec_t$
that are consistently estimated by $\widehat {\lambdavec}_i$ and $\widehat {\Fvec}_t$.\footnote{The matching method discussed here is also applicable to settings where the matching is based on variables other than the estimated factor structure. These include for example cross section and time series averages of the observable variables. See the appendix for a more general
treatment.}
We define 
\begin{align*}
     \Nset_i &= \Big\{ j \in \Nset \setminus \{i\} \, : \,\Big\| \widehat {\lambdavec}_i - \widehat {\lambdavec}_j \Big\|   \leq  \tau_{NT} \Big\} ,
     &
      \Tset_t &=   \Big\{ s \in \Tset \setminus \{t\} \, : \,   \Big\| \widehat {\Fvec}_t - \widehat {\Fvec}_s \Big\|  \leq  \upsilon_{NT} \Big\} ,
\end{align*}
for some bandwidth parameters $\tau_{NT}>0$ and $\upsilon_{NT} >0$. 
The debiased estimator of $\mu(x)$ is then given by
$$
\widetilde \mu(x) = \frac 1 {NT}  \sum_{i=1}^N \sum_{t=1}^T    \widetilde Y_{it}(x),
$$
with
\begin{align}
    \label{DefWidetildeY}
    \widetilde Y_{it}(x) = \left\{ \begin{array}{lll}
           Y_{it} \hfill \text{if $X_{it} =x$,}
           \\[10pt]
      \displaystyle \frac 1 {n_{it}} \sum_{j \in \Nset_i} \sum_{s \in \Tset_t}  \mathbbm{1}\{X_{is}=X_{jt}=X_{js}=x\} ( Y_{is} +  Y_{jt}  -  Y_{js}  )  \\
            \hfill \text{if $X_{it} \neq x$ and $n_{it}>0$,}
           \\[10pt]
          \frac 1 {  n(x)}  \sum_{(j,s) \in \mathbb{D}(x) } Y_{js} \hfill \text{if  $n_{it}=0$,}
    \end{array}
    \right.
\end{align}
where  $n_{it} :=  \sum_{j \in \Nset_i} \sum_{s \in \Tset_t}  \mathbbm{1}\{X_{is}=X_{jt}=X_{js}=x\}$.
Here, for $X_{it} \neq x$, we construct the counterfactual   $\widetilde Y_{it}(x) $ 
by averaging over all units $(j,s) \in \Nset_i \times \Tset_t$ that satisfy the constraint $X_{is}=X_{jt}=X_{js}=x$. 
Notice that if $X_{it} \neq x$ and $n_{it}=0$, then we cannot construct a suitable counterfactual by that method. In that case we assign
$ \widetilde Y_{it}(x)$ the average of the observations with $X_{js} = x$ to make sure that $\widetilde \mu(x) $ is always well-defined, but our assumption below guarantees that this rarely happens.

This estimator has similar debiasing properties to the nearest neighbor described above, but it is more tractable theoretically because it varies more smoothly with respect to the factors and loadings. 

Indeed,  $\widetilde \mu(x)$ can be written as 
$$\widetilde \mu(x) = \frac 1 {NT}  \sum_{i=1}^N \sum_{t=1}^T  \omega_{it} \, Y_{it},$$
       where the weights     $\omega_{it}$ are functions of 
        $\widehat {\lambdavec}_j$ and $\widehat {\Fvec}_s$ for all $j \in \Nset$ and $s \in \Tset$. 
To show that  $\widetilde \mu(x)$ is a consistent estimator of $\mu(x)$, we use the following assumption:
\begin{assumption}[Two-way Matching Estimator]
    \label{ass:Debias}
   There exists a sequence $\xi_{NT}>0$ such that $\xi_{NT}  \to 0$ as $N,T \to \infty$, and
    \begin{enumerate}[(a)]
        \item $ \frac 1 {NT}  \sum_{i=1}^N \sum_{t=1}^T    \mathbbm{1}\left\{ X_{it} \neq x \, \& \, n_{it}=0 \right\} = O_P \left(  \xi_{NT} \right)$.  
    
        \item  $Y_{it}$ is uniformly bounded over $i,t,N,T$.
        
        \item  $Y_{it}$ is independent across both $i$ and $t$, conditional on $\Xvec^{NT}$,  $\Avec^N$, $\Bvec^T$.
    
        \item The function $(\avec, \bvec)  \mapsto m(x, \avec, \bvec)$ is  at least twice continuously differentiable with uniformly bounded second derivatives.
        
        \item There exists $c>0$ such that  $\left\| \avec_1 - \avec_2 \right\| \leq c \left\| \lambdavec(\avec_1) -  \lambdavec(\avec_2) \right\|$ for all $\avec_1, \avec_2 \in \Aset$, and  $\left\| \bvec_1 - \bvec_2 \right\| \leq c \left\| \Fvec(\bvec_1) -  \Fvec(\bvec_2) \right\|$ for all $\bvec_1, \bvec_2 \in \Bset$.
              
       \item
       $ \frac 1 {N}  \sum_{i=1}^N \left(   \left\| \widehat{\lambdavec}_i - \lambdavec_i \right\|^2 
       +  \max_{j \in \Nset_i}  \left\| \widehat{\lambdavec}_j - \lambdavec_j \right\|^2 \right)= O_P \left( \xi_{NT} \right) $.
        \\
   $ \frac 1 {T}  \sum_{t=1}^T \left(   \left\| \widehat{\Fvec}_t - \Fvec_t \right\|^2 
       +  \max_{s \in \Tset_t}  \left\| \widehat{\Fvec}_s - \Fvec_s \right\|^2 \right)= O_P \left(  \xi_{NT} \right) $.
       
       \item
       $ \tau_{NT} ^2= O_P \left(   \xi_{NT} \right) $ and $\upsilon_{NT}^2 = O_P \left(   \xi_{NT} \right) $.
 
       \item  $
             \frac 1 {NT}  \sum_{i=1}^N \sum_{t=1}^T
       \Ep \left[  \omega^2_{it} \,\big| \, \Xvec^{NT}, \,  \Avec^N, \, \Bvec^T \right] 
                = O_P( NT \,  \xi_{NT}^2).
        $
 
       \item 
      Let $\Yvec^{NT}_{-(i,t),-(j,s)}$ be the outcome matrix  $\Yvec^{NT}$, but with $Y_{it}$ and $Y_{js}$ replace by zero (or some other non-random number),
      and all other outcomes unchanged. We assume 
       \begin{multline*}
               \frac 1 {(NT)^2}  \sum_{i,j=1}^N \sum_{t,s=1}^T 
               \mathbbm{1}\left\{ (i,t) \neq (j,s) \right\}
                 \Ep \bigg[  \Big|   \omega_{it}\left( \Yvec^{NT}_{-(i,t),-(j,s)} \right)  \, \omega_{js}\left( \Yvec^{NT}_{-(i,t),-(j,s)} \right) 
             \\      
                       -   \omega_{it}(\Yvec^{NT}) \, \omega_{js}(\Yvec^{NT})   \Big| 
                       \, \bigg| \,  \Xvec^{NT}, \,  \Avec^N, \, \Bvec^T \bigg] 
                =     O_P\left( \xi_{NT}^2 \right) .
         \end{multline*}
    \end{enumerate} 
\end{assumption}

\begin{remark}[Assumption~\ref{ass:Debias}]
\textnormal{
Part (i)  guarantees that $X_{it} \neq x$ and $n_{it}=0$ only happens for a small fraction of observations $(i,t)$.  We are therefore able to construct proper counterfactuals $ \widetilde Y_{it}(x)$ for most observations. Part (ii) is a boundedness condition that is standard in the matrix completion literature. Part (iii) is an independence condition that is convenient to simplify the derivations but can be generalized to weak correlation across both $i$ and $t$.  We use part (iv) to bound the error terms of the Taylor expansions for the bias. Part (v) imposes an injectivity condition.  The functions $\avec \mapsto \lambdavec(\avec)$ and $ \bvec \mapsto \Fvec(\bvec)$ need to be such that
        $\Avec_i$ and $\Bvec_t$ can be uniquely recovered from $\lambdavec_i = \lambdavec(\Avec_i)$ and  $\Fvec_t = \Fvec(\Bvec_t)$. A necessary condition is that the dimensions of $\lambdavec_i$ and $\Fvec_t$ are greater than or equal to the dimensions of $\Avec_i$ and $\Bvec_t$, respectively.  This holds in our factor structure approximation when let $R$ grow with the sample size, provided that the dimensions of $\Avec_i$ and $\Bvec_t$ are fixed. Part (vi) holds if  $ \widehat{\lambdavec}_i - \lambdavec_i$
       and $\widehat{\Fvec}_t - \Fvec_t $ are of order $N^{-1/2}$ and $T^{-1/2}$.
       We expect this assumption to be satisfied for rates $\xi_{NT} \gg \max(N^{-1},T^{-1})$.  
       The bandwidth parameters $ \tau_{NT}$ and $\upsilon_{NT}$ should not be chosen too large according to part (vii). For example, if we want to achieve a rate $   \xi_{NT} \ll \max(N^{-1/2},T^{-1/2})$, then 
       we need $ \tau_{NT} \ll \max(N^{-1/4},T^{-1/4})$ and $\upsilon_{NT} \ll \max(N^{-1/4},T^{-1/4})$.
Part (viii) requires that any given outcome $Y_{it}$ is not chosen too often with too high weight
        in the construction of the counterfactuals $  \widetilde Y_{js}(x) $.  Finally, part (ix) is a high-level assumption that could be justified by appropriate distributional assumptions
       on $X_{it}$, $\Avec_i$, $\Bvec_t$, and on the estimators $\widehat {\lambdavec}_i$ and $\widehat {\Fvec}_t$.        
       We prefer to present it as a high-level assumption, 
       because formally working out the distributional assumptions is quite cumbersome.
       Intuitively, if $n_{it}$ is sufficiently large, then changing  $\Yvec^{NT}$  to $ \Yvec^{NT}_{-(i,t),-(j,s)}$
       should not change the constructions of the counterfactual $\widehat Y_{it}(x)$ very much.
       If that is true for all $(i,t)$, then the weights $ \omega_{it}(\Yvec^{NT}) $ 
       should be very close to the weights $  \omega_{it}\left( \Yvec^{NT}_{-(i,t),-(j,s)} \right) $
       and the assumption is satisfied.}
\end{remark}

\begin{theorem}
      \label{ConvergenceRateMatching}
     Under Assumptions \ref{ass:model} and \ref{ass:Debias},
     \begin{align*}
           \widetilde \mu(x) -  \mu(x)    & = O_P \left( \xi_{NT} \right).
     \end{align*}
\end{theorem}

As discussed in the above remark, one can achieve rates $   \xi_{NT} \ll \max(N^{-1/2},T^{-1/2})$
for sufficiently regular data generating processes, and if
the bandwidth parameters  $ \tau_{NT}$ and $\upsilon_{NT}$ are chosen sufficiently small.
By contrast, the low-rank approximation bias  in $\widehat \mu(x)$ will usually prevent us from achieving such a
convergence rate for  $\widehat \mu(x)$. This finding is consistent with our Monte Carlo results in Section~\ref{sec:MC}, where
$ \widetilde \mu(x) $ is found to typically have much smaller bias than $\widehat \mu(x)$.

\section{Numerical Examples}\label{sec:examples}
\setcounter{equation}{0}

\subsection{Election day registration and voter turnout}
\label{sec:Election}

We illustrate the methods of the paper with an empirical application to the effect of allowing voter registration during the election day on voter turnout in the U.S. \citet{xu17}. Voting in the U.S. used to require registration prior to the election day in most states.   Registration increased the cost of voting and was considered as one possible reason for low turnout rates. In response, some states implemented Election Day Registration (EDR) laws that allowed eligible voters to register on election day when they arrive at the polling stations.  These laws were not  passed by all the states, and there was variation in the time of adoption across states. Thus, they were enacted by Maine, Minnesota and Wisconsin in 1976; Wyoming, Indiana and New Hampshire in 1994, and Connecticut in 2012. 

We use a dataset on the 24 presidential elections for 47 states between 1920 and 2012 collected by \citet{xu17}. It includes state-level information  about the turnout rate, $Y_{it}$, measured as the total ballots counted divided by voting-age population in state $i$ at election $t$, and a treatment indicator for EDR, $X_{it}$, that equals one if the state $i$ has an EDR law enacted at election $t$. Following \citet{xu17}, we exclude North Dakota where registration was never needed, and Alaska and Hawaii that were not states until 1959. Since there are only 9 states that are ever treated and the treatment started in the 1976 election, we focus on effects on the treated at the elections between 1976 and 2012. We estimate average treatment effects and quantile treatment effects at multiple quantile indices.

\begin{figure} 
\begin{center}
\includegraphics[height=.8\textwidth,width=.8\textwidth,angle=0]{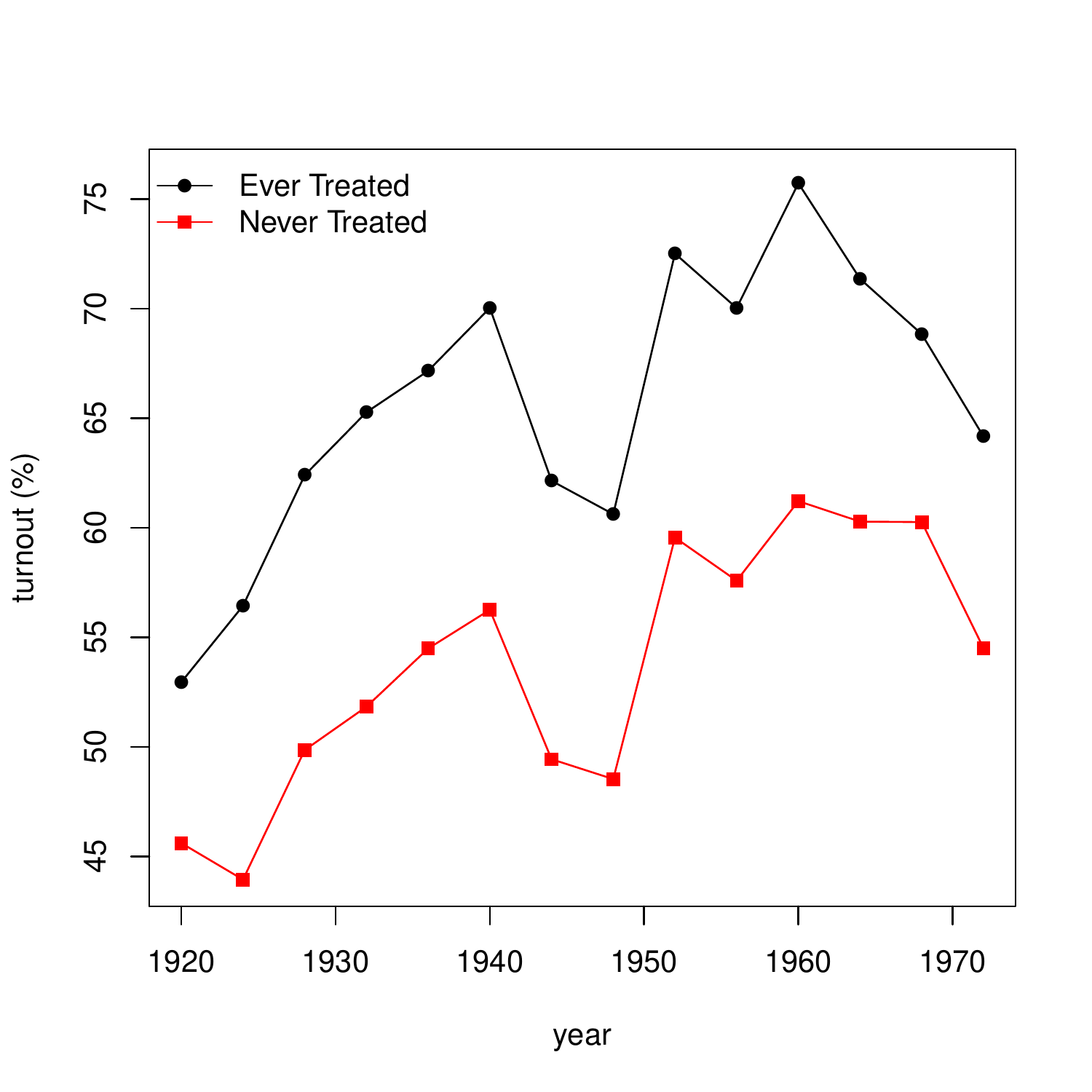}\caption{Pretrends in turnout rate}\label{fig:pretrends}
\end{center}
\end{figure}

Figure \ref{fig:pretrends} compares the average turnout of states that are ever treated with states that are never treated in elections prior to the first implementation of the EDR laws in 1976. It shows that  ever treated states have higher turnout rates on average than never treated states without the EDR treatment. We consider several methods to deal with this likely nonrandom assignment of EDR to estimate the ATTs for each election after 1976. First, we do a naive comparison of means between treated and nontreated states in each election (Dmeans). Second, we consider a difference-in-differences method that uses the nontreated states as controls at each election (DiD). In particular, we estimate the effects from a linear regression with state effects and election effects interacted with a EDR indicator. This method yields the ATT for each election under a parallel trend assumption between treated and nontreated states.\footnote{The DiD model is a special case our model with additive effects. In this case, it imposes that there are only additive state and election effects that are the same for both treatment levels.}    Third, we compute our estimator based on matrix completion methods without debiasing (MC) with additive state and election effects and the parameter $\rho$ such that the number of factors is $R = 6$. Fourth, we debias the MC estimates using the two-way matching method with 10 matches (TWM-10). Fifth, we consider  the simple matching method with 5 matches (SM-5). We choose the number of matches roughly based on the numerical simulations of Section \ref{sec:MC}.

\begin{figure}[ht]
        \begin{center}
            \includegraphics[height=.8\textwidth,width=.8\textwidth,angle=0]{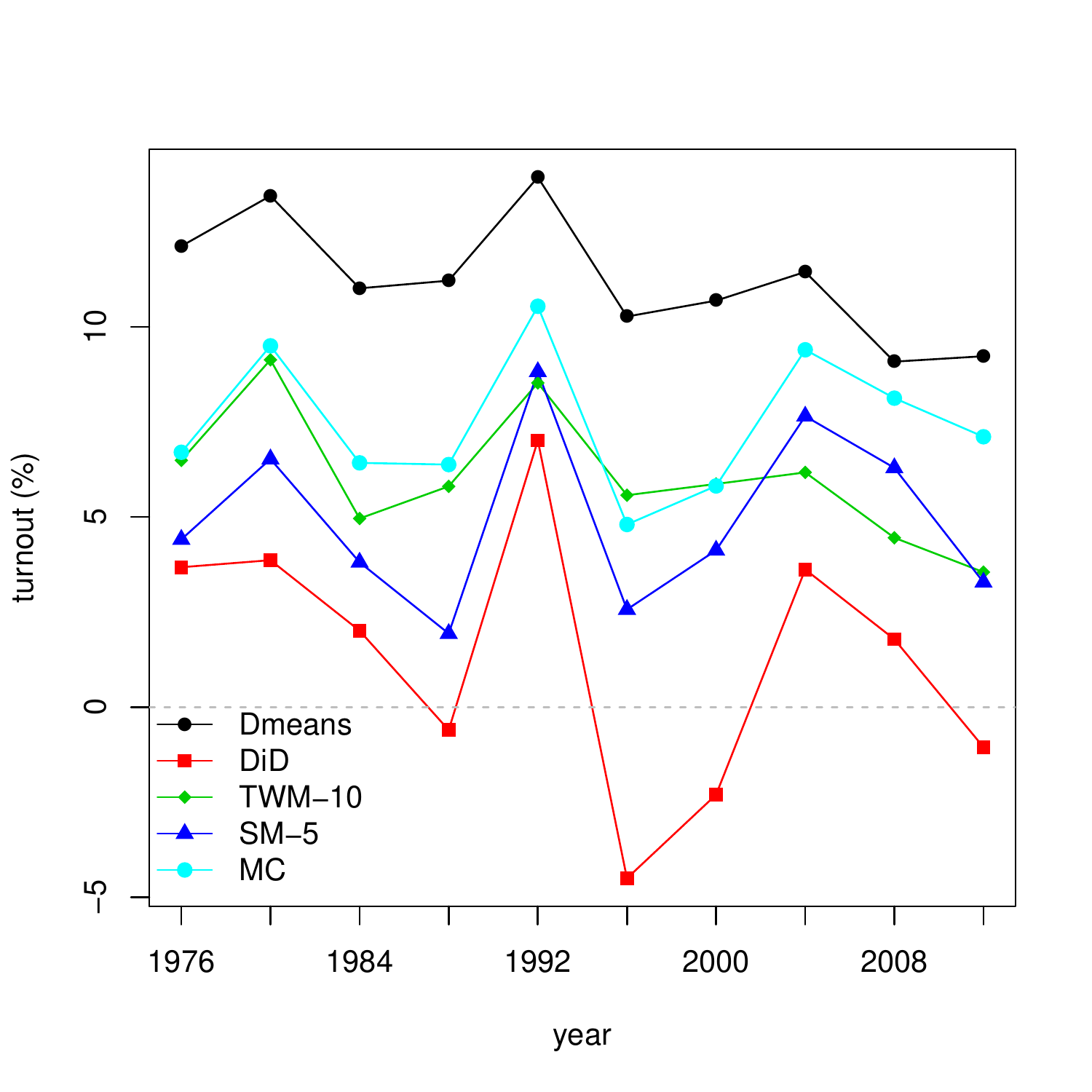}
            \caption{Average treatment effect on treated}\label{fig:ATT}
        \end{center}
    \end{figure}

Figure \ref{fig:ATT} reports the estimates of the ATT of EDR  at each election. The methods that account for possible nonrandom assignment of the EDR produce lower estimates of the effect than the naive comparison of means between treated and nontreated states. This finding agrees with the pre-EDR differences  found in fig.~\ref{fig:pretrends}.  MC, TWM-10 and SM-5 estimates are generally larger and more stable across elections than DiD estimates. According to TWM-10,  EDR laws increase voter turnout between 5 and 9\% depending on the election. This effect is an economically significant relative to 55\%, the average turnout rate for states without EDR. The estimates of the election-aggregated ATTs are 10.71\%, 0.67\%, 7.35\%,  5.56\%, and 4.87\% for Dmeans, DiD, MC, TWM-10, and SM-3, respectively. 

\begin{figure}[ht]
        \begin{center}
            \includegraphics[height=.8\textwidth,width=.8\textwidth,angle=0]{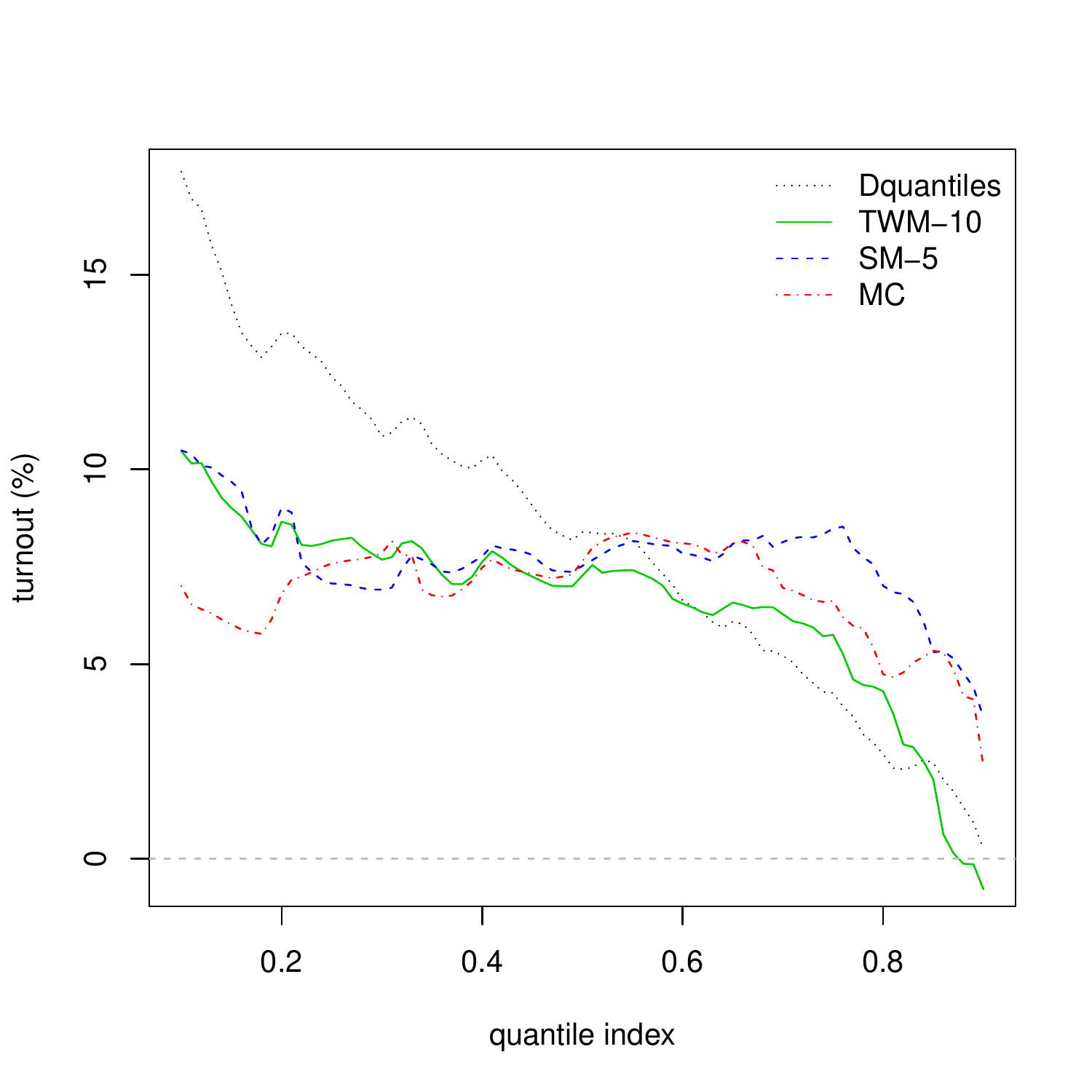}
            \caption{Time-averaged QTT}\label{fig:QTT}
        \end{center}
    \end{figure}

Figure \ref{fig:QTT} plots the estimates of the election-aggregated quantile treatment effect on the treated (QTT) of EDR as a function of the quantile index. We report estimates from four methods: a naive comparison of quantiles between treated and non-treated states (Dquantiles), our estimator based on matrix completion methods without debiasing (MC) with additive state and election effects and the parameter $\rho$ such that the number of factors is $R = 3$, two-way matching with 10 matches (TWM-10), and simple matching  with 5 matches (SM-5). The QTT is the difference of the quantiles  between the observed turnout for the treated observations and the corresponding potential turnout have they not been treated. The quantiles of the observed turnout are estimated using sample quantiles. The estimates of the quantiles of the potential outcomes are obtained by inverting the corresponding estimates of the distribution, which are obtained by our methods replacing $Y_{it}$ by the indicator $\indf(Y_{it} \leq y)$ and repeating the procedure over a grid of values of $y$ that includes the sample quantiles of observed turnout with indices  $\{.10, .11, \ldots, .98\}$.\footnote{We rearrange the estimates of the distribution to guarantee that they are increasing with respect to $y$ \citet{CFG10}.} Here, we find that the effect of EDR is decreasing across the distribution of turnout and ranges between 10 and 0\% according to TWM-10. EDR is therefore more effective at the bottom of the voter turnout distribution.  Comparing with the Dquantiles estimates, we find that the sign of the selection bias switches from positive to negative around the middle of the turnout distribution.

\subsection{Monte Carlo simulations}
\label{sec:MC}

To evaluate the performance of our methods in a  controlled synthetic environment, we generate potential outcomes from an additive linear model where  
$$
   Y_{it}(x) = x + g(A_i, B_t) + U_{it}(x), \quad x \in \{0,1\}, i \in \{1, \ldots, 30\}, t \in \{1, \ldots, 30\},
$$
${U}_{it}(x) \sim N(0,1/4)$ independently over $i$, $t$ and $x$, $A_i \sim U(0,1)$ independently over $i$,  $B_t \sim U(0,1)$ independently over $t$, ${U}_{it}(x)$, $A_j$ and $B_s$ are independent for all $i$, $t$, $j$ and $s$, and $g$ is the Gaussian kernel, i.e., 
\begin{align*}
    g(a,b) = \frac{1}{\sqrt{2\pi}\sigma}\exp\left(-\frac{(a-b)^2}{\sigma^2}\right). 
\end{align*}
This design is similar to that used in \citet{bordenave2020detection}, with kernel function specification from the numerical simulations in \citet{gh13}.\footnote{We find similar results in a multiplicative model where $Y_{it}(x) = (1 + x) g(A_i, B_t) + U_{it}(x)$. We omit these results for the sake of brevity.} The parameter $\sigma$ controls the decay of the singular values of $g$ and can be calibrated to make sure the singular values decay slowly. Smaller values for $\sigma$ lead to greater dispersion in the kernel function $(a,b) \mapsto g(a,b)$ and a slower singular value decay, hence can be interpreted as a measure of smoothness.\footnote{Smoothness here is specifically related to numerical smoothness, i.e. variability in the function within close neighbourhoods of its arguments.}  The assignment of $X_{it}$ that determines what potential outcomes are observed is similar to the election application. In particular, only observations for the first half of the units, $i \in \{1,\ldots, 15\}$, and the second half of the panel, $t \in \{ 15,\ldots,30 \}$,  may be treated. For these observations,  $X_{it}$ is related to the unobserved effects $(A_i,B_t)$ via  $X_{it} = \indf \{g(A_i,B_t)\geq c\}$, where $c$ is a constant calibrated to $\Pr(g(A_i,B_t)\geq c) =.5$.

\begin{table}[ht]
\caption{Results for  $\mu(0 \mid \{1\})$ }
\centering
\begin{tabular}{lcccc}
  \hline\hline
 &Bias & St. Dev.  & RMSE   \\%&&Bias & St. Dev.  & RMSE\\ 
  \hline
 Dmeans & 0.59 & 0.02 & 0.59 \\%& & 0.59 & 0.02 & 0.59  \\ 
  DiD & 0.70 & 0.03 & 0.70 \\%& & 0.69 & 0.03 & 0.70 \\ 
 MC & 0.74 & 0.02 & 0.74 \\%& & 0.74 & 0.02 & 0.74 \\ 
 TWM-1 & 0.03 & 0.14 & 0.14 \\%& & 0.02 & 0.13 & 0.14 \\ 
  TWM-5 & 0.03 & 0.11 & 0.12 \\%& & 0.03 & 0.11 & 0.11 \\ 
 TWM-10 & 0.04 & 0.10 & 0.11 \\%& & 0.04 & 0.10 & 0.11  \\ 
 TWM-30 & 0.07 & 0.09 & 0.12 \\%& & 0.06 & 0.09 & 0.11 \\ 
  SM-1 & 0.12 & 0.10 & 0.16 \\%& & 0.12 & 0.11 & 0.16 \\ 
  SM-5 & 0.15 & 0.07 & 0.17 \\%& & 0.15 & 0.07 & 0.17\\ 
  SM-10 & 0.19 & 0.06 & 0.20 \\%& & 0.19 & 0.06 & 0.20\\ 
  SM-30 & 0.31 & 0.05 & 0.31\\% & & 0.31 & 0.05 & 0.31\\ 
    \hline\hline
    \multicolumn{4}{l}{\footnotesize{Notes: based on $1,000$ simulations}}
\end{tabular}
 \label{table:CASF}
\end{table}

\begin{figure}[ht]
\centering
  \includegraphics[width=0.32\linewidth, height = 6cm]{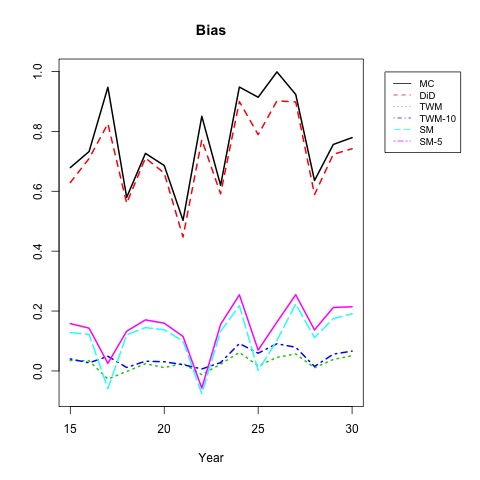}
  \includegraphics[width=0.32\linewidth, height = 6cm]{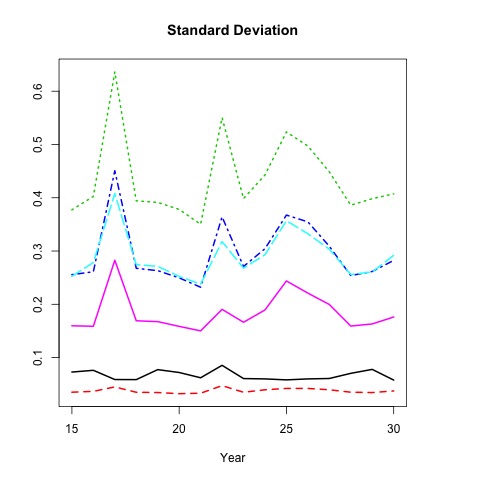}
  \includegraphics[width=0.32\linewidth, height = 6cm]{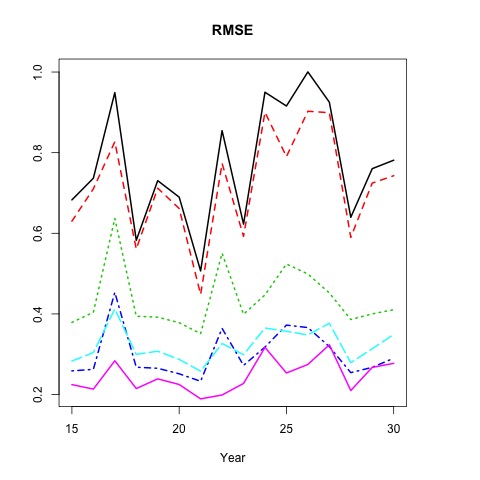}
\caption{Results for  $t \mapsto \mu_t(0 \,|\, \{1\})$.}
\label{fig:sim_linear}
\end{figure}

We apply similar methods to Section \ref{sec:Election} to estimate the CASFs  $\mu_t(0 \mid \{1\}),$ $t \in \{ 15,\ldots,30 \}$, and $\mu(0 \mid \{1\})$ using  the observed variables $X_{it}$ and $Y_{it} = Y_{it}(X_{it})$.  Thus, we consider Dmeans, DiD, MC without additive effects and with the parameter $\rho$ such that $R=5$, and multiple versions of TWM and SM with the number of matches equal to $1$, $5$, $10$, and $30$. For each method, we compute the bias, standard deviation and rmse from $1,000$ simulations. Across the simulations, we redraw the values of $U_{it}(x)$ and hold $A_i$, $B_t$ and $X_{it}$ fixed.  Table \ref{table:CASF} reports the results for the time-aggregated CASF,  $\mu(0 \mid \{1\})$, and Figure \ref{fig:sim_linear} plots the results for the CASF, $\mu_t(0 \mid \{1\})$, as a function of $t$. The results show that Dmeans, DiD and MC are severely biased relative to their standard deviations. All the matching estimators reduce bias and rmse, despite of increasing dispersion. As one would expect, increasing the number of matches reduces the variability of the matching estimators but increases their biases.  The number of matches that minimizes the rmse is larger for the TWM than for the SM.  Overall, these small-sample findings agree with the asymptotic results of  Sections \ref{sec:mc} and \ref{sec:debias}.

\section*{Acknowledgements}
This paper was prepared for the Econometrics Journal Special Session on ``Econometrics of Panel Data'' at the  Royal Economic Society 2019 Annual Conference in Warwick University. We thank  the editor Jaap Abbring, two anonymous referees, Shuowen Chen, and the participants of this conference  and the 25$^{\text{th}}$ International Panel Data Conference for comments. This research was
supported by the Economic and Social Research Council through the ESRC Centre for
Microdata Methods and Practice grant RES-589-28-0001, and by  the
European Research Council grants ERC-2014-CoG-646917-ROMIA and
ERC-2018-CoG-819086-PANEDA.

    \section*{Appendix A: Proofs of Results}
    \renewcommand{\theequation}{A.\arabic{equation}}
    \renewcommand{\thesection}{A}
    \renewcommand{\theassumption}{A.\arabic{assumption}}
    \renewcommand{\thelemma}{A.\arabic{lemma}}
    \renewcommand{\thetheorem}{A.\arabic{theorem}}
    \renewcommand{\theproposition}{A.\arabic{proposition}}
    \setcounter{equation}{0}

    \medskip

We start with a preliminary result that relates the nuclear norm of  $\Gammavec^{\infty}(x)$ with the sum of the singular values of the function $(\avec, \bvec) \mapsto m(x,\avec,\bvec) $.  This link will be useful to bound the approximation error of $\widehat \Gammavec(x)$. 
We define $$ \left\| m(x, \cdot , \cdot)  \right\|_* :=  \sum_{j=1}^\infty s_j(x).$$

\begin{lemma}
    \label{lemma:NuclearNormBound}
    Let Assumptions~\ref{ass:SamplingAB} and \ref{ass:Smoothness} hold.
    Then, as $N,T \rightarrow \infty$, 
     \begin{align*}
           \left\|  \Gammavec^{\infty}(x) \right\|_1 \, \leq \, \sqrt{NT}  \,   \left\| m(x, \cdot , \cdot)  \right\|_* + o_P(\sqrt{NT}) 
           = O_P(\sqrt{NT}).
     \end{align*}
\end{lemma}
\medskip
 Lemma \ref{lemma:NuclearNormBound} implies that  $   \left\|  \Gammavec^{\infty}(x) \right\|_1$ grows with $N$ and $T$ at  the same rate as any low-rank matrix $ \Mvec$ with elements that are of order one with bounded second moments such that  $   \left\|  \Mvec \right\|_1 \leq  \sqrt{{\rm rank}(\Mvec)}   \left\|   \Mvec \right\|_2 
= \sqrt{{\rm rank}( \ \Mvec) \,  \sum_{i=1}^N \sum_{t=1}^T  M_{it}^2}   = O_P(\sqrt{NT})$. 
This result will be useful for  the proofs of  Lemma~\ref{lemma:consistency1} and of Theorem~\ref{th:MAIN1}.
The proof of Lemma~\ref{lemma:NuclearNormBound} is provided {at the end of the appendix.}

The following technical lemma provides the key step in the proof of Lemma~\ref{lemma:consistency1} in the main text.
\begin{lemma}
      \label{lemma:BoundGammaObserved}
      Under Assumptions~\ref{ass:SamplingAB} and \ref{ass:Smoothness}, 
     \begin{align*}
            \frac 1 {  n(x)}  \sum_{(i,t) \in \mathbb{D}(x) }
 \left( \widehat \Gamma_{it}(x)  - \Gamma^{\infty}_{it}(x) \right)^2 \leq  \frac{2 \, \rho \,  \|  \Gammavec^{\infty}(x) \|_1 } {n(x)}
  - \frac 2 {n(x)} \sum_{(i,t) \in \mathbb{D}(x) } \Gamma^{\infty}_{it}(x) E_{it},
     \end{align*}
for all $\rho \geq \| \Evec(x) \|_\infty$.
\end{lemma}
Notice that Lemma~\ref{lemma:BoundGammaObserved} is a non-stochastic finite sample result,
which only requires that $  E_{it}(x) $  and $\widehat \Gammavec(x)$ are as defined in  \eqref{DefE} and \eqref{DefEstimator}.   
The proof of Lemma~\ref{lemma:BoundGammaObserved} is provided {at the end of the appendix.} We are now ready to provide the proof of the lemma in the main text.
 
\textbf{Proof of Lemma~\ref{lemma:consistency1}:}
    
The definition of $E_{it}(x)$ in  \eqref{DefE} guarantees that\\ $  \Ep\left[E_{it}(x) \mid \Avec^N, \Bvec^T, \Xvec^{NT} \right] = 0$,
and Assumption~\ref{ass:SamplingE} furthermore guarantees that $E_{it}(x)$ is independent across $i$ and $t$
and has a finite fourth moment, conditional on $\Xvec^{NT}$, $\Avec^N$ and $\Bvec^T$.
Furthermore, $ \Gamma^{\infty}_{it}(x) = m(x, \Avec_i, \Bvec_t)$ only depends on $ \Avec^N$ and $\Bvec^T$. 
We therefore find
\begin{align*}
  &   \Ep\left[ \left. \left(  \frac 1 {n(x)} \sum_{(i,t) \in \mathbb{D}(x) } \Gamma^{\infty}_{it}(x) \, E_{it} \right)^2 \,  \right| \, \Avec^N, \Bvec^T, \Xvec^{NT} \right]  
\\
     & \qquad =    \frac 1 {n^2(x)}   \sum_{(i,t) \in \mathbb{D}(x) } 
     \left[ \Gamma^{\infty}_{it}(x) \right]^2 \,  \Ep\left[  \left. E_{it}^2 \,  \right| \, \Avec^N, \Bvec^T, \Xvec^{NT} \right] 
 \\
    &  \qquad  \leq      \frac {b^{1/2}} {n^2(x)}   \sum_{(i,t) \in \mathbb{D}(x) } 
     \left[ \Gamma^{\infty}_{it}(x) \right]^2 = O_P(1/n(x)),
\end{align*} 
where $b$ is the constant from Assumption~\ref{ass:SamplingE}. 
 From this we conclude that
\begin{align}
     \frac 1 {n(x)} \sum_{(i,t) \in \mathbb{D}(x) } \Gamma^{\infty}_{it}(x) \, E_{it} 
     = O_P\left(  \frac 1 {n^{1/2}(x)}  \right)
     = o_P(1).
     \label{GammaInfE}
\end{align}
Next, applying Assumption~\ref{ass:SamplingE} and Theorem 2 in \cite{latala05} we find
      \begin{align*}
            \Ep\left[  \| \Evec(x) \|_\infty  \mid \Avec^N, \Bvec^T, \Xvec^{NT}  \right] 
           & \leq C \Bigg\{
                 \max_t \sqrt{ \sum_i    \Ep\left[E_{it}(x)^2 \mid \Avec^N, \Bvec^T, \Xvec^{NT}  \right] }                
        \\ & \qquad \qquad
             +  \max_i \sqrt{ \sum_t    \Ep\left[E_{it}(x)^2 \mid \Avec^N, \Bvec^T, \Xvec^{NT}  \right] }  
        \\ &\qquad \qquad \qquad
              +    \left(  \sum_{i,t}    \Ep\left[E_{it}(x)^4 \mid \Avec^N, \Bvec^T, \Xvec^{NT}  \right]    \right)^{1/4} 
            \Bigg\} 
           \\
           &\leq  C \,  b^{1/4} \, \left\{   \sqrt{   N}  +  \sqrt{   T} + n(x)^{1/4}    \right\} = O_P\left( \sqrt{N+T} \right),
      \end{align*}
      where $C$ is a universal constant. We therefore have
$
   \| \Evec(x) \|_\infty =  O_P\left( \sqrt{N+T} \right)
$,
and since we assume that $\rho = \rho_{NT}$ satisfies $\rho_{NT} /  \sqrt{N+T} \rightarrow \infty$ we conclude that
\begin{align*}
     \rho_{NT} \geq \| \Evec(x) \|_\infty
\end{align*}
with probability approaching one. We can therefore apply Lemma~\ref{lemma:BoundGammaObserved} 
to find that, with probability approaching one, we have
     \begin{align*}
            \frac 1 {  n(x)}  \sum_{(i,t) \in \mathbb{D}(x) }
 \left( \widehat \Gamma_{it}(x)  - \Gamma^{\infty}_{it}(x) \right)^2 
 & \leq  \frac{2 \, \rho_{NT} \,  \|  \Gammavec^{\infty}(x) \|_1 } {n(x)}
  - \frac 2 {n(x)} \sum_{(i,t) \in \mathbb{D}(x) } \Gamma^{\infty}_{it}(x) E_{it}
\\
 & =   \frac{2 \, \rho_{NT} \, O_P(\sqrt{NT}) } {n(x)} + o_P(1) 
\\ 
  &= o_P(1) ,
     \end{align*}
where we applied \eqref{GammaInfE} and Lemma~\ref{lemma:NuclearNormBound}, as well as the condition $\rho_{NT} \sqrt{NT} / n(x) \rightarrow 0$.

\hfill$\square$\\

In the following consider a generic reduced form parameter 
\begin{align}\label{eq:rfpar}
    \nu_0(x) = \frac 1 {NT}  \sum_{i=1}^N \sum_{t=1}^T   W_{it}(x) \, \Gamma^{\infty}_{it}(x),
\end{align}
with corresponding estimator
\begin{align}\label{nuHAT}
\widehat \nu(x) = \frac 1 {NT}  \sum_{i=1}^N \sum_{t=1}^T   W_{it}(x) \, \widehat \Gamma_{it}(x),
\end{align}
where $W_{it}(x) $ are given weights.

The following proposition provides a finite-sample non-stochastic bound for the error of this reduced form estimator. 
\begin{proposition}
      \label{prop:BoundMain}
    Let the Assumptions~\ref{ass:SamplingAB}, \ref{ass:Smoothness} and \ref{ass:SamplingE} hold.  
   Let $P_{it}(x)$ be non-zero real numbers for all $(i,t) \in \Nset \times \Tset$.
    Define
    \begin{align*}
         V_{it}(x) &:=  \frac{    W_{it}(x)  \, P_{it}^{-1}(x) (D_{it}(x) - P_{it}(x))      } {\frac 1 {NT} \sum_{i=1}^N \sum_{t=1}^T W_{it}(x)^2 P_{it}^{-1}(x)} ,
    \\
         c_1 &:=  \frac{ 1 - \frac 1 {NT}  \sum_{i=1}^N \sum_{t=1}^T W_{it}(x) P_{it}^{-1}(x)  V_{it}(x) }
            {\frac 1 {NT} \sum_{i=1}^N \sum_{t=1}^T W_{it}(x)^2 P_{it}^{-1}(x)} ,
   \\    
        c_2 &:=   \frac {1} {NT}  \sum_{i=1}^N \sum_{t=1}^T    V_{it}(x)  \,  \Gamma^{\infty}_{it}(x)  ,
    \\
        c_3 &:=      \frac{2 \, \rho }{c_1 \, NT} \|  \Gammavec^{\infty}(x) \|_1  
  -      \frac 2 {c_1 \, NT}  \sum_{(i,t) \in \mathbb{D}(x)}   E_{it}(x) \,  \Gamma^{\infty}_{it}(x) + \left(  \frac{c_2} { c_1}  \right)^2,
  \\ 
       c_4 &:=  \sqrt{c_3}   +  \frac{|c_2|}{c_1}   ,
    \end{align*}  
     and let
     $\Vvec(x)$ be  the $N \times T$ matrix with elements $V_{it}(x)$.
     If $c_1 > 0$ 
    and    
   $\rho > \| \Evec(x) \|_\infty +c_4  \| \Vvec(x) \|_\infty $,
  then       
   \begin{align*}
         \left| \widehat \nu(x) - \nu_0(x)   \right|  \leq c_4.
    \end{align*}
\end{proposition}      
\medskip
The proof of Proposition~\ref{prop:BoundMain} is provided {at the end of the appendix.}
 Proposition~\ref{prop:BoundMain} is the key step required for the proof of 
Theorem~\ref{th:MAIN1}. However, before proving this main text result we want to provide an informal remark on the usefulness of
 Proposition~\ref{prop:BoundMain} more generally.

\begin{remark}[Consistency of $\widehat \nu(x)$]
\textnormal{
Proposition~\ref{prop:BoundMain} holds for all $P_{it}(x)  \in \mathbb{R} \setminus \{0\}$, but for the proposition to be useful in showing consistency of $  \widehat \nu(x)$
we need to choose $P_{it}(x)$ such that
$c_2$ and $ \| \Vvec(x) \|_\infty$ are  not too large. The easiest way to guarantee this
is to consider $X_{it}$ to be random and weakly correlated across both $i$ and $t$, and to define $P_{it}(x) $ as the propensity score, that is,
\begin{align*}
     P_{it}(x) =   \Pr \left( X_{it} = x \mid \Avec^N, \Bvec^T  \right),
\end{align*}
which is assumed to be positive and not too small --- e.g.\ we need that 
$$q:= \left[ \frac 1 {NT} \sum_{i=1}^N \sum_{t=1}^T W_{it}(x)^2 P_{it}^{-1}(x) \right]^{-1}$$
converges to some positive constant.
Then $  V_{it}(x)$ has mean zero, analogous to $E_{it}(x)$, and 
\begin{align*}
    c_1 &=  q  +  O_P(1/\sqrt{NT})  ,
      \\
    c_2 &=    O_P(1/\sqrt{NT})      
    \\
      c_3 &=     \frac{2 \, \rho }{q\, NT} \|  \Gammavec^{\infty}(x) \|_1   +  O_P(1/\sqrt{NT}) ,
    \\
     c_4 &=   \sqrt{  \frac{2 \, \rho }{ q \, NT} \|  \Gammavec^{\infty}(x) \|_1  } + \text{smaller order terms} .
\end{align*}   
Thus, if, like in Lemma~\ref{lemma:consistency1},  $\rho = \rho_{NT}$ such that $\rho_{NT} /  \sqrt{N+T} \rightarrow \infty$ and $\rho_{NT} / \sqrt{NT}   \rightarrow 0$
      as $N,T \rightarrow \infty$,  then 
\begin{align*}
    \widehat \nu(x) = \nu_0(x) + o_P(1). 
\end{align*}      
The following proof formalizes this heuristic argument for the case that $W_{it}(x)=1$.}
\end{remark}

\textbf{Proof of Theorem~\ref{th:MAIN1}:}
Let $W_{it}(x) = 1$,  
and let $ \nu_0(x)$ and $\widehat \nu(x)$ be as defined in  \eqref{eq:rfpar} and \eqref{nuHAT} above.
We then have
\begin{align}
\mu(x) &=  \nu_0(x)   ,
 \nonumber \\
\widehat \mu(x) &= \widehat \nu(x) 
   +    \frac 1 {NT} \sum_{(i,t) \in \mathbb{D}(x) }  E_{it}(x)
   -  \frac 1 {NT}  \sum_{(i,t) \in \mathbb{D}(x) } \left[  \widehat \Gamma_{it}(x) -  \Gamma^{\infty}_{it}(x) \right] .
   \label{HatMuProof}
\end{align}
We drop all the arguments $x$ in the rest of this proof.
We want to apply Proposition~\ref{prop:BoundMain} with  $  P_{it} =   \Pr \left( X_{it} = x \mid \Avec^N, \Bvec^T  \right) >0$.
Let $ G_{it}  =    P_{it}^{-1} (D_{it} - P_{it}) $ be as defined in  Theorem~\ref{th:MAIN1},
and also define $q:= \left[ \frac 1 {NT} \sum_{i=1}^N \sum_{t=1}^T   P_{it}^{-1} \right]^{-1}$.
Since $P_{it} \in [0,1]$ we also have $q \in [0,1]$, and the theorem assumes that $q^{-1} = O_P(1)$.
Using Lemma~\ref{lemma:NuclearNormBound} we know that
$ \|  \Gammavec^{\infty} \|_1 = O_P(\sqrt{NT})$,
and we have already found that 
$  \sum_{(i,t) \in \mathbb{D} } \Gamma^{\infty}_{it} \, E_{it} = O_P\left( n^{1/2}  \right)$
in \eqref{GammaInfE} above.
Using this together the other assumptions in the theorem we find that
\begin{align*}
    V_{it}  &=  q  \, G_{it}  
    \\
   c_1 &= q  \left( 1 -   \frac q {NT}  \sum_{i=1}^N \sum_{t=1}^T   P_{it}^{-1}   \, G_{it}  \right)
    = q \,  [1-o_P(1)]  ,
   \\    
        c_2 &=   \frac {q} {NT}  \sum_{i=1}^N \sum_{t=1}^T     G_{it}   \,  \Gamma^{\infty}_{it}  = o_P(1) ,
    \\
        c_3 &=      \frac{2 \, \rho \, O_P(\sqrt{NT}) }{c_1 \, NT}  
  -      \frac {O_P\left( n^{1/2}  \right)} {c_1 \, NT}   + \left(  \frac{c_2} { c_1}  \right)^2 
  =  o_P(1) ,
  \\ 
       c_4 &=  \sqrt{c_3}   +  \frac{|c_2|}{c_1} =  o_P(1)  .
\end{align*}  
We furthermore have
\begin{align*}
  \| \Vvec \|_\infty &= q \,  \| \Gvec \|_\infty
 =  O_P(1) \,  O_P(\sqrt{N+T}) = O_P(\sqrt{N+T}) .
\end{align*}
In the proof of Lemma~\ref{lemma:consistency1} we already argued that  $  \| \Evec \|_\infty =  O_P\left( \sqrt{N+T} \right)$.
Since we assume that $\rho = \rho_{NT}$ satisfies $\rho_{NT} /  \sqrt{N+T} \rightarrow \infty$ we conclude that
\begin{align*}
    \rho > \| \Evec \|_\infty +c_4  \| \Vvec \|_\infty  
\end{align*}
with probability approach one. We can therefore apply Proposition~\ref{prop:BoundMain} 
to find that with probability approach one we have
   \begin{align*}
         \left| \widehat \nu - \nu_0   \right|  \leq c_4 = o_P(1) .
    \end{align*}
We have thus shown that     $\widehat \nu = \nu_0 + o_P(1) $.

Furthermore, analogous to the result in \eqref{GammaInfE}
we can show that
$  \sum_{(i,t) \in \mathbb{D} }   E_{it} = O_P\left( n^{1/2}  \right)$,
and we therefore have $  \frac 1 {NT} \sum_{(i,t) \in \mathbb{D} }  E_{it} = o_P(1)$.
Finally, applying Lemma~\ref{lemma:consistency1}  we have
 Next, from we know that
     \begin{align*}
         \left[    \frac 1 {  n} \sum_{(i,t) \in \mathbb{D} } \left(  \widehat \Gamma_{it} -  \Gamma^{\infty}_{it} \right) \right]^2
          &\leq     
            \frac 1 {  n}  \sum_{(i,t) \in \mathbb{D} }
 \left( \widehat \Gamma_{it}  - \Gamma^{\infty}_{it} \right)^2  
 = o_P(1),
     \end{align*}
and therefore $ \frac 1 {NT}  \sum_{(i,t) \in \mathbb{D}(x) } \left[  \widehat \Gamma_{it}(x) -  \Gamma^{\infty}_{it}(x) \right] = o_P(1)$.
Plugging those result into \eqref{HatMuProof} we find $  \widehat \mu(x) = \mu(x)  + o_P(1)$. 

 \hfill$\square$\\

In this section we present and prove a more general version of Theorem~\ref{ConvergenceRateMatching}.
Let $\boldsymbol{\phi}_i = \boldsymbol{\phi}(x,\Avec_i)$ and  $\boldsymbol{\psi}_t = \boldsymbol{\psi}(x,\Bvec_t)$ be transformations of $\Avec_i$ 
and $\Bvec_t$. Let $\widehat {\boldsymbol{\phi}}_i$ and $\widehat {\boldsymbol{\psi}}_t$ be corresponding estimators.
In the main text we presented the special case where $\widehat {\boldsymbol{\phi}}_i$ and $\widehat {\boldsymbol{\psi}}_t$
were equal to the factor loadings and factors obtained from $\widehat \Gammavec(x)$, but many other choices of 
$\widehat {\boldsymbol{\phi}}_i$ and $\widehat {\boldsymbol{\psi}}_t$ are conceivable. 
We again define 
\begin{align*}
     \Nset_i &= \left\{ j \in \Nset \setminus \{i\} \, : \,\left\| \widehat {\boldsymbol{\phi}}_i - \widehat {\boldsymbol{\phi}}_j \right\|   \leq  \tau_{NT} \right\} ,
     &
      \Tset_t &=   \left\{ s \in \Tset \setminus \{t\} \, : \,   \left\| \widehat {\boldsymbol{\psi}}_t - \widehat {\boldsymbol{\psi}}_s \right\|  \leq  \upsilon_{NT} \right\} ,
\end{align*}
for some bandwidth parameters $\tau_{NT}>0$ and $\upsilon_{NT} >0$. 
A debiased estimator of the reduced form parameter in \eqref{eq:rfpar} is given by
$$
\widetilde \nu(x) = \frac 1 {NT}  \sum_{i=1}^N \sum_{t=1}^T   W_{it}(x) \, \widetilde Y_{it}(x),
$$
where $\widetilde Y_{it}(x)$ is defined as in \eqref{DefWidetildeY}. In the main text we only discussed the special case
$  W_{it}(x)=1$. We can write $\widetilde \nu(x)$
 as 
$$\widetilde \nu(x) = \frac 1 {NT}  \sum_{i=1}^N \sum_{t=1}^T  \omega_{it} \, Y_{it},$$
       where the weights     $\omega_{it}$ are functions of 
        $\widehat {\boldsymbol{\phi}}_j$ and $\widehat {\boldsymbol{\psi}}_s$ for all $j \in \Nset$ and $s \in \Tset$. 
Assumption~\ref{ass:Debias} in the main text is generalized as follows.        
 
\begin{assumption} 
    \label{ass:DebiasGeneral}
   There exists a sequence $\xi_{NT}>0$ such that $\xi_{NT}   \to 0$ as $N,T \to \infty$, and
    \begin{enumerate}[(a)]
        \item $ \frac 1 {NT}  \sum_{i=1}^N \sum_{t=1}^T   W_{it}(x) \, \mathbbm{1}\left\{ X_{it} \neq x \, \& \, n_{it}=0 \right\} = O_P \left(  \xi_{NT} \right)$.
    
        \item  $Y_{it}$ and  $W_{it}(x)$ are uniformly bounded over $i,t,N,T$.
        
        \item  $Y_{it}$ is independent across both $i$ and $t$, conditional on $\Xvec^{NT}$,  $\Avec^N$, $\Bvec^T$.
    
        \item The function $(\avec, \bvec)  \mapsto m(x, \avec, \bvec)$ is twice continuously differentiable with uniformly bounded second derivatives.
        
        \item There exists $c>0$ such that  $\left\| \avec_1 - \avec_2 \right\| \leq c \left\| \boldsymbol{\phi}(\avec_1) -  \boldsymbol{\phi}(\avec_2) \right\|$ for all $\avec_1, \avec_2 \in \Aset$, and  $\left\| \bvec_1 - \bvec_2 \right\| \leq c \left\| \boldsymbol{\psi}(\bvec_1) -  \boldsymbol{\psi}(\bvec_2) \right\|$ for all $\bvec_1, \bvec_2 \in \Bset$.
              
       \item
       $ \frac 1 {N}  \sum_{i=1}^N \left(   \left\| \widehat{\boldsymbol{\phi}}_i - \boldsymbol{\phi}_i \right\|^2 
       +  \max_{j \in \Nset_i}  \left\| \widehat{\boldsymbol{\phi}}_j - \boldsymbol{\phi}_j \right\|^2 \right)= O_P \left( \xi_{NT} \right) $.
        \\
   $ \frac 1 {T}  \sum_{t=1}^T \left(   \left\| \widehat{\boldsymbol{\psi}}_t - \boldsymbol{\psi}_t \right\|^2 
       +  \max_{s \in \Tset_t}  \left\| \widehat{\boldsymbol{\psi}}_s - \boldsymbol{\psi}_s \right\|^2 \right)= O_P \left(  \xi_{NT} \right) $.
       
       \item
       $ \tau_{NT} ^2= O_P \left(   \xi_{NT} \right) $ and $\upsilon_{NT}^2 = O_P \left(   \xi_{NT} \right) $.
 
       \item  $
             \frac 1 {NT}  \sum_{i=1}^N \sum_{t=1}^T
       \Ep \left[  \omega^2_{it} \,\big| \, \Xvec^{NT}, \,  \Avec^N, \, \Bvec^T \right] 
                = O_P( NT \,  \xi_{NT}^2).
        $
 
       \item 
      Let $\Yvec^{NT}_{-(i,t),-(j,s)}$ be the outcome matrix  $\Yvec^{NT}$, but with $Y_{it}$ and $Y_{js}$ replace by zero (or some other non-random number),
      and all other outcomes unchanged. We assume 
       \begin{multline*}
               \frac 1 {(NT)^2}  \sum_{i,j=1}^N \sum_{t,s=1}^T 
               \mathbbm{1}\left\{ (i,t) \neq (j,s) \right\}
                 \Ep \bigg[  \Big|   \omega_{it}\left( \Yvec^{NT}_{-(i,t),-(j,s)} \right)  \, \omega_{js}\left( \Yvec^{NT}_{-(i,t),-(j,s)} \right) 
             \\      
                       -   \omega_{it}(\Yvec^{NT}) \, \omega_{js}(\Yvec^{NT})   \Big| 
                       \, \bigg| \,  \Xvec^{NT}, \,  \Avec^N, \, \Bvec^T \bigg] 
                =     O_P\left( \xi_{NT}^2 \right) .
         \end{multline*}
    \end{enumerate} 
\end{assumption}

The generalized version of Theorem~\ref{ConvergenceRateMatching} is given in the following.
\begin{theorem}
      \label{ConvergenceRateMatchingGeneral}
     Under Assumptions \ref{ass:model} and \ref{ass:DebiasGeneral},
     \begin{align*}
           \widetilde \nu(x) -  \nu_0(x)    & = O_P \left( \xi_{NT} \right).
     \end{align*}
\end{theorem}

 \textbf{Proof of Theorem~\ref{ConvergenceRateMatchingGeneral} (containing Theorem~\ref{ConvergenceRateMatching} as a special case)}

Define $m_{it}(x) := m(x, \Avec_i, \Bvec_t) $.
We  decompose 
\begin{align}
    \widetilde \nu(x) -  \nu_0(x)  &=  e_0(x) +  e_1(x) + e_2(x), 
    \label{Def_e12}
\end{align}
where
\begin{align*}
     e_0(x) &=  \frac 1 {NT}  \sum_{i=1}^N \sum_{t=1}^T   W_{it}(x) \, \mathbbm{1}\left\{ X_{it} \neq x \, \& \, n_{it}=0 \right\}  \left[ m_{it}(X_{it}) - m_{it}(x) \right] ,
\end{align*}
and
\begin{align*}
     e_1(x)  &:= 
        \frac 1 {NT}  \sum_{i=1}^N \sum_{t=1}^T  \mathbbm{1}\left\{ X_{it} \neq x \, \& \, n_{it}>0 \right\}  \,   W_{it}(x) \, e_{1,it}(x)  ,
     \\
     e_{1,it}(x) &:=
      \frac{ \sum_{j \in \Nset_i} \sum_{s \in \Tset_t}  \mathbbm{1}\{X_{is}=X_{jt}=X_{js}=x\} \left[ m_{is}(x)  +  m_{jt}(x)   -  m_{js}(x)  - m_{it}(x)  \right] }
       { \sum_{j \in \Nset_i} \sum_{s \in \Tset_t}  \mathbbm{1}\{X_{is}=X_{jt}=X_{js}=x\}},
\end{align*}
and
\begin{align*}
     e_2(x)  &:= \frac 1 {NT}  \sum_{i=1}^N \sum_{t=1}^T  \omega_{it} \, E_{it}   ,
\end{align*}
In the following we consider $ e_0(x) $, $e_1(x)$, $e_2(x)$ separately.

\bigskip
\noindent
\# \underline{Bound on $ e_0(x)$:}
  Assumption~\ref{ass:DebiasGeneral}(i) and (ii)  guarantee that
  \begin{align}
    \left| e_0(x) \right| &\leq
   \left( \max_{it} \left| m_{it}(X_{it}) - m_{it}(x) \right|  \right)
      \frac 1 {NT}  \sum_{i=1}^N \sum_{t=1}^T   W_{it}(x) \, \mathbbm{1}\left\{ X_{it} \neq x \, \& \, n_{it}=0 \right\} 
           \nonumber  \\
          &= O_P \left( \xi_{NT} \right)  .
         \label{bound_e0}
 \end{align}
    
\bigskip
\noindent
\# \underline{Bound on $ e_1(x)$:}
    Assumption~\ref{ass:DebiasGeneral}(iv)  guarantees that there exists a constant $b>0$ such that
 \begin{align*}
   &  \left| 
     m(x, \avec, \bvec) -  m(x, \Avec_i, \Bvec_t)  -  (\avec - \Avec_i)' \, \frac{\partial m(x, \Avec_i, \Bvec_t)} {\partial \Avec_i}
     -  (\bvec - \Bvec_t)' \, \frac{\partial m(x, \Avec_i, \Bvec_t)} {\partial \Bvec_t}
     \right|
     \\
    & \qquad \qquad \qquad \qquad \qquad \qquad \qquad  \qquad \qquad \qquad   \qquad \qquad 
      \leq b  \left(  \left\| \avec - \Avec_i \right\|^2 +  \left\| \bvec - \Bvec_t \right\|^2 \right) .
 \end{align*}
 Using this we find that
 \begin{align*}
    m_{is}(x)  +  m_{jt}(x)   -  m_{js}(x)  - m_{it}(x) 
    \leq 2 \, b \, \left( \left\| \Avec_i - \Avec_j \right\|^2 + \left\|  \Bvec_t  -  \Bvec_s  \right\|^2 \right) ,
 \end{align*}
 and therefore
 \begin{align*}
       \left|   e_{1,it}(x)  \right| &\leq
      \frac{2 \, b \, \sum_{j \in \Nset_i} \sum_{s \in \Tset_t}  \mathbbm{1}\{X_{is}=X_{jt}=X_{js}=x\}
      \left( \left\| \Avec_i - \Avec_j \right\|^2 + \left\|  \Bvec_t  -  \Bvec_s  \right\|^2 \right)
        }
       { \sum_{j \in \Nset_i} \sum_{s \in \Tset_t}  \mathbbm{1}\{X_{is}=X_{jt}=X_{js}=x\}}
     \\
      &\leq  2 \, b \left( \max_{j \in \Nset_i}  \left\| \Avec_i - \Avec_j \right\|^2
                            + \max_{s \in \Tset_t}  \left\|  \Bvec_t  -  \Bvec_s  \right\|^2 \right) .
 \end{align*}
 We thus find
 \begin{align*}
       | e_1(x) |  &\leq 
       2 \, b   \Big( \max_{ij} \left| W_{it}(x) \right| \Big)
         \Bigg(  \frac 1 {N}  \sum_{i=1}^N   \max_{j \in \Nset_i}  \left\| \Avec_i - \Avec_j \right\|^2
          +
           \frac 1 {T}    \sum_{t=1}^T    \max_{s \in \Tset_t}  \left\|  \Bvec_t  -  \Bvec_s  \right\|^2
           \Bigg)
      \\
       &\leq     
         2 \, b \, c \,  \Big( \max_{ij} \left| W_{it}(x) \right| \Big)
         \Bigg(  \frac 1 {N}  \sum_{i=1}^N   \max_{j \in \Nset_i}  \left\| \boldsymbol{\phi}(\Avec_i) - \boldsymbol{\phi}(\Avec_j) \right\|^2
         \\
           & + \frac 1 {T}    \sum_{t=1}^T    \max_{s \in \Tset_t}  \left\|   \boldsymbol{\psi}( \Bvec_t )  -  \boldsymbol{\psi}(  \Bvec_s )  \right\|^2
           \Bigg) 
      \\
       &=    
         2 \, b \, c \,  \Big( \max_{ij} \left| W_{it}(x) \right| \Big)
         \Bigg( \frac 1 {N}  \sum_{i=1}^N   \max_{j \in \Nset_i}  \left\| \boldsymbol{\phi}_i - \boldsymbol{\phi}_j \right\|^2
          +
           \frac 1 {T}    \sum_{t=1}^T    \max_{s \in \Tset_t}  \left\|   \boldsymbol{\psi}_t  -  \boldsymbol{\psi}_s   \right\|^2
           \Bigg) .
 \end{align*}
 Using the triangle inequality, the definition of $\Nset_i$, and the general inequality $(x_1+x_2+x_3)^2 \leq 3 (x_1^2 + x_2^2 + x_3^2)$, for $x_1,x_2,x_3 \in \mathbb{R}$,  we have
 \begin{align*}
    \max_{j \in \Nset_i}    \left\| \boldsymbol{\phi}_i  - \boldsymbol{\phi}_j \right\|^2
     & \leq   \max_{j \in \Nset_i}  
      \left( \left\| \widehat{\boldsymbol{\phi}}_i - \widehat{\boldsymbol{\phi}}_j \right\|
      +  \left\|   \widehat{\boldsymbol{\phi}}_i - \boldsymbol{\phi}_i  \right\|
        + \left\|   \widehat{\boldsymbol{\phi}}_j - \boldsymbol{\phi}_j  \right\| \right)^2
\\
 & \leq   \max_{j \in \Nset_i}  
      \left( \tau_{NT}  
      +  \left\|   \widehat{\boldsymbol{\phi}}_i - \boldsymbol{\phi}_i  \right\|
        + \left\|   \widehat{\boldsymbol{\phi}}_j - \boldsymbol{\phi}_j  \right\| \right)^2      
 \\
  &\leq   3 \tau_{NT} ^2  + 3  \left\|   \widehat{\boldsymbol{\phi}}_i - \boldsymbol{\phi}_i  \right\|^2
   +  3  \max_{j \in \Nset_i}    \left\|   \widehat{\boldsymbol{\phi}}_j - \boldsymbol{\phi}_j  \right\|^2 .
 \end{align*}
 Analogously we find
 \begin{align*}
   \max_{s \in \Tset_t}  \left\|   \boldsymbol{\psi}_t  -  \boldsymbol{\psi}_s \right\|^2
     &\leq   3 \upsilon_{NT}^2
      + 3  \left\|   \widehat{\boldsymbol{\psi}}_t - \boldsymbol{\psi}_t  \right\|^2
   +  3  \max_{s \in \Tset_t}    \left\|   \widehat{\boldsymbol{\psi}}_s - \boldsymbol{\psi}_s  \right\|^2 .
 \end{align*}
 We thus obtain
  \begin{align}
       \left| e_1(x) \right|  &\leq 
       6 \, b \, c \,  \left( \max_{ij} \left| W_{it}(x) \right| \right)
         \Bigg\{
         \tau_{NT} ^2 +\upsilon_{NT}^2
      \nonumber \\ & \qquad     \qquad \qquad \qquad \qquad \qquad 
        + \frac 1 {N}  \sum_{i=1}^N \left(   \left\| \widehat{\boldsymbol{\phi}}_i - \boldsymbol{\phi}_i \right\|^2 
       +  \max_{j \in \Nset_i}  \left\| \widehat{\boldsymbol{\phi}}_j - \boldsymbol{\phi}_j \right\|^2 \right)
   \nonumber  \\ & \qquad     \qquad \qquad \qquad \qquad \qquad 
        + \frac 1 {T}  \sum_{t=1}^T \left(   \left\| \widehat{\boldsymbol{\psi}}_t - \boldsymbol{\psi}_t \right\|^2 
       +  \max_{s \in \Tset_t}  \left\| \widehat{\boldsymbol{\psi}}_s - \boldsymbol{\psi}_s \right\|^2 \right)       \Bigg\}
           \nonumber  \\
          &= O_P \left( \xi_{NT} \right)  .
         \label{bound_e1}
 \end{align}

\bigskip
\noindent
\# \underline{Bound on $ e_2(x)$:}
We have
\begin{align*}
 \left[ e_2(x) \right]^2   &= \frac 1 {(NT)^2}  \sum_{i,j=1}^N \sum_{t,s=1}^T  \omega_{it}(\Yvec^{NT}) \,  \omega_{js}(\Yvec^{NT})  \, E_{it} \, E_{js}  = T_0 +  T_1 + T_2  ,
\end{align*}
where
\begin{align*}
    T_0 &:=  \frac 1 {NT}  \sum_{i=1}^N \sum_{t=1}^T  \omega^2_{it}(\Yvec^{NT})  \, E^2_{it}  ,
   \\
   T_1 &:= \frac 1 {(NT)^2}   \sum_{i,j=1}^N \sum_{t,s=1}^T  \mathbbm{1}\left\{ (i,t) \neq (j,s) \right\}
   \\ & \qquad  \qquad   \times 
  \left[ \omega_{it}(\Yvec^{NT}) \, \omega_{js}(\Yvec^{NT}) 
                       -  \omega_{it}\left( \Yvec^{NT}_{-(i,t),-(j,s)} \right)  \, \omega_{js}\left( \Yvec^{NT}_{-(i,t),-(j,s)} \right)  \right]  \, E_{it} \, E_{js}  ,
  \\                     
   T_2 &:=    \frac 1 {(NT)^2}   \sum_{i,j=1}^N \sum_{t,s=1}^T  \mathbbm{1}\left\{ (i,t) \neq (j,s) \right\}
            \omega_{it}\left( \Yvec^{NT}_{-(i,t),-(j,s)} \right)  \, \omega_{js}\left( \Yvec^{NT}_{-(i,t),-(j,s)} \right)  \, E_{it} \, E_{js}           .           
\end{align*}
We have
\begin{align*}
     \Ep \left[ T_0 \, \Big| \,  \Xvec^{NT}, \,  \Avec^N, \, \Bvec^T \right]  &\leq 
      \left(  \max_{i,t} \left| E_{it} \right| \right)^2
     \frac 1 {(NT)^2}  \sum_{i=1}^N \sum_{t=1}^T
       \Ep \left[  \omega^2_{it}\left( \Yvec^{NT}  \right) \Big| \,   \Xvec^{NT}, \,  \Avec^N, \, \Bvec^T \right] 
         \\
         &= O_P(\xi^2_{NT})  ,             
\end{align*}
and
\begin{align*}
   & \left|  \Ep \left[ T_1 \, \Big| \,  \Xvec^{NT}, \,  \Avec^N, \, \Bvec^T \right]  \right|
    \\
    &\leq
   \left(  \max_{i,t} \left| E_{it} \right| \right)^2     
      \frac 1 {(NT)^2}    \sum_{i,j=1}^N \sum_{t,s=1}^T  \mathbbm{1}\left\{ (i,t) \neq (j,s) \right\} 
      \\ &   \times
  \Ep \left[  \left| \omega_{it}(\Yvec^{NT}) \, \omega_{js}(\Yvec^{NT}) 
                       -  \omega_{it}\left( \Yvec^{NT}_{-(i,t),-(j,s)} \right)  \, \omega_{js}\left( \Yvec^{NT}_{-(i,t),-(j,s)} \right)  \right| 
                       \, \Big| \,  \Xvec^{NT}, \,  \Avec^N, \, \Bvec^T \right] 
         \\
         &= O_P(\xi^2_{NT})  .      
\end{align*}
where we used that $Y_{it}$ (and thus  $E_{it}$) is uniformly bounded, together with Assumption~\ref{ass:DebiasGeneral}(viii) and (ix).
Next, for $(i,t) \neq (j,s)$ we 
 \begin{align*}
     & \Ep \left[  \omega_{it}\left( \Yvec^{NT}_{-(i,t),-(j,s)} \right)  \, \omega_{js}\left( \Yvec^{NT}_{-(i,t),-(j,s)} \right)   \, E_{it} \, E_{js} \, \Big| \, \Yvec^{NT}_{-(i,t),-(j,s)}, \, \Xvec^{NT}, \,  \Avec^N, \, \Bvec^T \right]
     \\
     &=    \omega_{it}\left( \Yvec^{NT}_{-(i,t),-(j,s)} \right)  \, \omega_{js}\left( \Yvec^{NT}_{-(i,t),-(j,s)} \right) 
      \Ep \left[   E_{it} \, E_{js} \, \Big| \, \Yvec^{NT}_{-(i,t),-(j,s)}, \, \Xvec^{NT}, \,  \Avec^N, \, \Bvec^T \right]
   \\
    &=      \omega_{it}\left( \Yvec^{NT}_{-(i,t),-(j,s)} \right)  \, \omega_{js}\left( \Yvec^{NT}_{-(i,t),-(j,s)} \right) 
    \\ & \qquad \qquad
      \Ep \left[   E_{it}   \, \Big| \, \Yvec^{NT}_{-(i,t),-(j,s)}, \, \Xvec^{NT}, \,  \Avec^N, \, \Bvec^T \right]
        \Ep \left[   E_{js} \, \Big| \, \Yvec^{NT}_{-(i,t),-(j,s)}, \, \Xvec^{NT}, \,  \Avec^N, \, \Bvec^T \right]
   \\
    &= 0,
 \end{align*}
 where we used $  \Ep \left[ E_{it} \mid  \, \Xvec^{NT} \, ,  \Avec^N, \, \Bvec^T   \right]  = 0$ together with the assumption that $Y_{it}$ (and thus $E_{it}$) is independent across both $i$ and $t$, conditional on $\Xvec^{NT}$,  $\Avec^N$, $\Bvec^T$. By the law of iterated expectations the last display result also implies that  for $(i,t) \neq (j,s)$ we have
 \begin{align*}
       \Ep \left[  \omega_{it}\left( \Yvec^{NT}_{-(i,t),-(j,s)} \right)  \, \omega_{js}\left( \Yvec^{NT}_{-(i,t),-(j,s)} \right)   \, E_{it} \, E_{js} \, \Big| \,  \Xvec^{NT}, \,  \Avec^N, \, \Bvec^T \right] &= 0.
 \end{align*}
Using this we obtain that
\begin{align*}
     \Ep \left[ T_2 \, \Big| \,  \Xvec^{NT}, \,  \Avec^N, \, \Bvec^T \right]  &=   0 .
\end{align*}
Combining those results on $T_0$, $T_1$, $T_2$ we obtain
\begin{align*}
       \Ep \left\{   \left[ e_2(x) \right]^2 \, \Big| \,  \Xvec^{NT}, \,  \Avec^N, \, \Bvec^T \right\}  =  O_P(\xi^2_{NT}),
\end{align*}
which implies $e_2 = O_P(\xi_{NT})$.
Together with \eqref{Def_e12}, \eqref{bound_e0}, and  \eqref{bound_e1} this gives the statement of the theorem.

\hfill$\square$\\

\textbf{Proof of Lemma~\ref{lemma:NuclearNormBound}:}
    Let $\uvec_j(x)$ be the $N$-vector with elements $u_j(x,\Avec_i)$,
    and let $\vvec_j(x)$ be the $T$-vector with elements $v_j(x,\Bvec_t)$. Then we have
    $ \Gammavec^{\infty}(x) = \sum_{j=1}^\infty \, s_j(x)  \, \uvec_j(x) \, \vvec_j^\T(x)$, and therefore
    \begin{align*}
         &  \left\|  \Gammavec^{\infty}(x) \right\|_1 
            \leq  \sum_{j=1}^\infty \, s_j(x)  \,  \left\|  \uvec_j(x) \right\| \, \left\| \vvec_j(x) \right\|
           \\
            &=  \sqrt{NT} \,   \sum_{j=1}^\infty  \, s_j(x)  \,  \sqrt{\frac 1 N \sum_{i=1}^N [u_j(x,\Avec_i)]^2 }
             \,  \sqrt{\frac 1 T \sum_{t=1}^T [v_j(x,\Bvec_t)]^2 }
           \\
           &\leq     \sqrt{NT} \,   \sum_{j=1}^\infty  \, s_j(x)  \, 
           \left( 1 + \frac{ \frac 1 {N} \sum_{i=1}^N [u_j(x,\Avec_i)]^2 - 1 } 2  \right)
              \left( 1 + \frac{ \frac 1 {T} \sum_{t=1}^T [v_j(x,\Bvec_t)]^2 - 1 } 2  \right) 
           \\
           &=       \sqrt{NT} \,   \sum_{j=1}^\infty  \, s_j(x) +  \sqrt{NT} \, R_{NT}
           \\
           &=  \sqrt{NT}  \,   \left\| m(x, \cdot , \cdot)  \right\|_*  +  \sqrt{NT} \, R_{NT} ,
    \end{align*}
    where for the second inequality we used that $\sqrt{z} \leq 1 + \frac{z-1} 2$, for all $z \geq 0$, and we defined $  R_{NT}  = \frac 1 {NT} \sum_{i=1}^N \sum_{t=1}^T  r_{it}$,
    with
    \begin{align*}
         r_{it} =   \sum_{j=1}^\infty  \, s_j(x) \left\{ \frac{ [u_j(x,\Avec_i)]^2 +  [v_j(x,\Bvec_t)]^2} 4 +  \frac{ [u_j(x,\Avec_i)]^2   [v_j(x,\Bvec_t)]^2} 4 - \frac 3 4 \right\} .
    \end{align*}
    Assumption~\ref{ass:Smoothness} guarantees that $ [u_j(x,\Avec_i)]^2$ and $[v_j(x,\Bvec_t)]^2$ have mean equal to one,
    which implies that $ r_{it}$ has mean zero. 
    Assumption~\ref{ass:SamplingAB} and the WLLN therefore guarantees that $R_{NT} = o_P(1)$.
    We have thus shown that
    $   \left\|  \Gammavec^{\infty}(x) \right\|_1 \, \leq \, \sqrt{NT}  \,   \left\| m(x, \cdot , \cdot)  \right\|_* + o_P(\sqrt{NT}) $,
    and since $ \left\| m(x, \cdot , \cdot)  \right\|_* $ is finite and non-random  we also have $ \left\|  \Gammavec^{\infty}(x) \right\|_1 = O_P(\sqrt{NT})$.

\hfill$\square$\\

\textbf{Proof of Lemma~\ref{lemma:BoundGammaObserved}}
    The nuclear norm (or trace norm) can be defined by 
    \begin{align}
            \|  \Gammavec \|_1 &=
             \max_{\left\{  \Mvec \in \Rset^{N \times T} \, : \, \|\Mvec\|_\infty \leq 1  \right\}} 
             \underbrace{ {\rm Tr}\left(  \Mvec'  \Gammavec \right) }_{\text{\scriptsize $\displaystyle  = \sum_{i=1}^N \sum_{t=1}^T M_{it} \Gamma_{it}$}}
             .
             \label{DefTraceNorm}
    \end{align}
    Our assumption $\rho \geq \| \Evec(x) \|_\infty$ guarantees that
    a possible choice in this maximization is $\Mvec = \rho^{-1} \Evec(x)$, and we therefore have
    \begin{align*}
          \rho \, \|  \Gammavec  \|_1  \geq  \sum_{i=1}^N \sum_{t=1}^T   D_{it}(x) \, E_{it}(x) \;  \Gamma_{it}  .
    \end{align*}
    Using this and the model  $  Y_{it} = \Gamma^{\infty}_{it}(x) + E_{it}(x)$,   for $X_{it}=x$, we find that
    \begin{align*}
        &  Q_{NT}(  \Gammavec,\rho,x)
        \\
        &=  \frac 1 2 \sum_{i=1}^N \sum_{t=1}^T 
  D_{it}(x)
 \left(Y_{it} - \Gamma_{it} \right)^2  + \rho \|  \Gammavec \|_1     
        \\
           &\geq  \frac 1 2 \sum_{i=1}^N \sum_{t=1}^T 
  D_{it}(x)
 \left( \Gamma^{\infty}_{it}(x) + E_{it}(x) - \Gamma_{it} \right)^2  + \sum_{i=1}^N \sum_{t=1}^T   D_{it}(x) \, E_{it}(x) \;  \Gamma_{it}   
         \\ 
      &=    \frac 1 2 \sum_{i=1}^N \sum_{t=1}^T 
  D_{it}(x)  \left( \Gamma^{\infty}_{it}(x)  - \Gamma_{it} \right)^2 
  +  \sum_{i=1}^N \sum_{t=1}^T 
   D_{it}(x) \, \Gamma^{\infty}_{it}(x) E_{it}(x)
  \\
  &+    \frac 1 2 \sum_{i=1}^N \sum_{t=1}^T   D_{it}(x) \, E_{it}^2(x) .
     \end{align*}
  By definition we have
    \begin{align*}
         Q_{NT}(\widehat \Gammavec(x),\rho,x) \leq  Q_{NT}( \Gammavec^{\infty}(x),\rho,x) = 
            \frac 1 2 \sum_{i=1}^N \sum_{t=1}^T   D_{it}(x) \, E_{it}^2(x) +  \rho \|  \Gammavec^{\infty}(x) \|_1
    \end{align*}
 Combining the results in the last two displays gives the statement of the lemma.

\hfill$\square$\\
 
 \textbf{Proof of Proposition~\ref{prop:BoundMain}}
 In this proof we drop the argument $x$ everywhere, and we define $\theta = NT \nu$ and $\theta_0 = NT \nu_0$. 
Define the $NT$-vectors
 $\gammavec = {\rm vec}(\Gammavec)$,  $\gammavec^{\infty} = {\rm vec}(\Gammavec^{\infty})$,
  $\wvec = {\rm vec}(W_{it} : i \in \Nset, t \in \Tset)$,
  $\dvec = {\rm vec}(D_{it} : i \in \Nset, t \in \Tset)$, and
 $\pvec = {\rm vec}(P_{it} : i \in \Nset, t \in \Tset)$.
 Then, $\diag(\pvec)$ is an $NT \times NT$ diagonal matrix.
For $\rho>0$ and $\theta \in \mathbb{R}$ we define
\begin{align*}
     L_{NT}(  \theta,\rho) =   \min_{ \left\{ \Gammavec \in \mathbb{R}^{N \times T} \, : \, \theta = \wvec' \gammavec \right\} }  Q_{NT}(  \Gammavec,\rho) ,
\end{align*}
which is the profile objective function that minimizes $Q_{NT}(  \Gammavec,\rho) $ over almost all parameters $ \Gammavec$, only keeping 
our parameter of interest fixed at $ \theta = \wvec' \gammavec = \sum_{i=1}^N \sum_{t=1}^T  W_{it} \Gamma_{it}$. Our goal is to show that the minimizing value
\begin{align*}
   \widehat \theta := \argmin_{\theta \in \mathbb{R}} L_{NT}(  \theta,\rho)  = \sum_{i=1}^N \sum_{t=1}^T  W_{it} \widehat \Gamma_{it}
\end{align*}
is close to $ \theta := \wvec' \gammavec^{\infty} =   \sum_{i=1}^N \sum_{t=1}^T  W_{it} \Gamma^{\infty}_{it}$.
Using the definition of $Q_{NT}(  \Gammavec,\rho) $  and $Y_{it} = \Gamma^{\infty}_{it} + E_{it}$,   for $D_{it}=1$, we find that
\begin{align}
     L_{NT}(  \theta,\rho) \leq     Q_{NT}(  \Gammavec^{\infty},\rho) =    \frac 1 2 \sum_{i=1}^N \sum_{t=1}^T   D_{it} \, E_{it}^2
       + \rho \|  \Gammavec^{\infty} \|_1 .
     \label{LowerBoundLtrue}  
\end{align}
If for a given value of $\theta = \wvec' \gammavec$  we have that the matrix $\Mvec(\theta)$ with elements 
 $M_{it}(\theta) :=  D_{it}  \,   E_{it}  -  \frac{ \wvec' (\gammavec - \gammavec^{\infty}) } { \wvec' \diag(\pvec)^{-1}  \wvec}  \frac{(D_{it} - P_{it}) W_{it}} {P_{it}}  $ satisfies
$\| \Mvec(\theta) \|_\infty \leq \rho$,
then by the definition of $\| \cdot \|_1$ in \eqref{DefTraceNorm} we have $ \rho \|  \Gammavec \|_1    \leq {\rm Tr}(\Gammavec' \Mvec(\theta)) =  \sum_{i=1}^N \sum_{t=1}^T M_{it}(\theta) \Gamma_{it} $. 
Using this and  $  Y_{it} = \Gamma^{\infty}_{it} + E_{it}$,   for $D_{it}=1$, we find that
{\small
\begin{align*}
 &   Q_{NT}(  \Gammavec,\rho)
 =  \frac 1 2 \sum_{i=1}^N \sum_{t=1}^T   D_{it}   \left(Y_{it} - \Gamma_{it} \right)^2 
        + \rho \|  \Gammavec \|_1     
 \\
           &\geq  \frac 1 2 \sum_{i=1}^N \sum_{t=1}^T  D_{it}
 \left[ \left( \Gamma^{\infty}_{it}  - \Gamma_{it} \right) + E_{it} \right]^2 
  +   \sum_{i=1}^N \sum_{t=1}^T  \left\{  D_{it}  \,   E_{it}  -   \frac{ [(\gammavec - \gammavec^{\infty})' \,  \wvec] } { \wvec' \diag(\pvec)^{-1}  \wvec}   \frac{(D_{it} - P_{it}) W_{it}} {P_{it}}  \right\}  \, \Gamma_{it} 
 \\
        &=  
        \underbrace{  \frac 1 2 \sum_{i=1}^N \sum_{t=1}^T  D_{it}
 \left( \Gamma_{it} - \Gamma^{\infty}_{it}   \right)^2  
   -   \frac{ [(\gammavec - \gammavec^{\infty})' \,  \wvec] } { \wvec' \diag(\pvec)^{-1}  \wvec}    \sum_{i=1}^N \sum_{t=1}^T    \frac{(D_{it} - P_{it}) W_{it}} {P_{it}}   \, \left( \Gamma_{it} - \Gamma^{\infty}_{it} \right) 
   }_{=: Q^{({\rm low},1)}_{NT}(  \Gammavec)}
  \\ & \qquad \qquad \qquad \qquad \qquad \qquad \qquad \qquad \qquad \qquad
+  \underbrace{
    \sum_{i=1}^N \sum_{t=1}^T    M_{it}(\theta) \,  \Gamma^{\infty}_{it}  
  +    \frac 1 2 \sum_{i=1}^N \sum_{t=1}^T   D_{it} \, E_{it}^2
    }_{=: Q^{({\rm low},2)}_{NT} } ,
\end{align*}}
where in the last step we added and subtracted $\sum_{i=1}^N \sum_{t=1}^T    M_{it}(\theta) \,  \Gamma^{\infty}_{it}  $,
and we multiplied out $ \left[ \left( \Gamma^{\infty}_{it}  - \Gamma_{it} \right) + E_{it} \right]^2 $, which leads to some simplifications.
Notice that $D_{it} \, E_{it} = E_{it}$ by construction of $E_{it}$, so that some occurrences of $D_{it}$ above could be dropped, but we find it clearer to keep
track of $D_{it}$ explicitly here.

Next, we
define the $NT \times NT$ idempotent matrices ${\bf P}=  \frac{ \diag(\pvec)^{-1}  \wvec \, \wvec'} { \wvec' \diag(\pvec)^{-1}  \wvec}$
and  ${\bf R} =  {\bf I}_{NT} - {\bf P}$. 
We then have
{\small
\begin{align*}
 &  Q^{({\rm low},1)}_{NT}(  \Gammavec)
  \\
   &= \frac 1 2 (\gammavec - \gammavec^{\infty})' \, {\rm diag}(\dvec) \,(\gammavec - \gammavec^{\infty})
      - \frac{ [(\gammavec - \gammavec^{\infty})' \,  \wvec] } { \wvec' \diag(\pvec)^{-1}  \wvec}    \left[  \wvec' \, \diag(\pvec)^{-1} \, {\rm diag}(\dvec - \pvec)   \,(\gammavec - \gammavec^{\infty}) \right]
 \\  
   &= \frac 1 2 (\gammavec - \gammavec^{\infty})' \left( {\bf P}'  + {\bf R}' \right)
   {\rm diag}(\dvec)     \left(  {\bf P} +  {\bf R} \right)
      (\gammavec - \gammavec^{\infty})\\
      &-  (\gammavec - \gammavec^{\infty})'   {\bf P}'   {\rm diag}(\dvec - \pvec)  \left( {\bf P}  + {\bf R} \right) (\gammavec - \gammavec^{\infty}) 
 \\
  &=       \frac 1 2 (\gammavec - \gammavec^{\infty})'   {\bf R}'   {\rm diag}(\dvec)   {\bf R}     (\gammavec - \gammavec^{\infty})
   +    \frac 1 2 (\gammavec - \gammavec^{\infty})'   {\bf P}'   {\rm diag}\left( 2 \pvec -  \dvec    \right)   {\bf P}     (\gammavec - \gammavec^{\infty}) ,
 \\
  &=       \frac 1 2 (\gammavec - \gammavec^{\infty})'   {\bf R}'   {\rm diag}(\dvec)   {\bf R}     (\gammavec - \gammavec^{\infty})
   \\
   &+   \frac 1 2 (\gammavec - \gammavec^{\infty})'   {\bf P}'   {\rm diag}\left( \pvec -  \dvec    \right)   {\bf P}     (\gammavec - \gammavec^{\infty}) 
   + \frac 1 2 \frac{  \left[ (\gammavec - \gammavec^{\infty})'   \wvec \right]^2 } { { \wvec' \diag(\pvec)^{-1}  \wvec}}
\end{align*}  
}
where all the ``mixed terms'' (that involve both $ {\bf P} $ and $ {\bf R}$) cancel because
we have ${\bf P}' \, {\rm diag}(\pvec) \, {\bf R} = 0$,
and in the last step we used that $ {\bf P}'  \, {\rm diag}\left( \pvec     \right) \,  {\bf P} =\frac{  \wvec \,  \wvec' } { { \wvec' \diag(\pvec)^{-1}  \wvec}}$.
We have
\begin{align*}
     \min_{ \left\{ \Gammavec \in \mathbb{R}^{N \times T} \, : \, \theta = \wvec' \gammavec \right\} }  
         (\gammavec - \gammavec^{\infty})' \,  {\bf R}'  \, {\rm diag}(\dvec) \,  {\bf R}  \,   (\gammavec - \gammavec^{\infty})
     &= 0,
\end{align*} 
because $\gammavec^* = {\bf R} \gammavec^{\infty} +  \theta \frac{ \diag(\pvec)^{-1} w } {w' \diag(\pvec)^{-1} w}$ is a possible choice in the minimization problem,
which satisfies $w' \gammavec^* = \theta$ and $  {\bf R}  \,   (\gammavec^* - \gammavec^{\infty}) = 0$. We therefore have
\begin{align*}
     &  \min_{ \left\{ \Gammavec \in \mathbb{R}^{N \times T} \, : \, \theta = \wvec' \gammavec \right\} }  
        Q^{({\rm low},1)}_{NT}(  \Gammavec)
      \\
      &=    
    \frac 1 2 \left( \theta- \theta_0 \right)^2  
   \left( \frac 1 { \wvec' \diag(\pvec)^{-1}  \wvec}
     +  \frac{\wvec'   \diag(\pvec)^{-1}   {\rm diag}\left( \pvec -  \dvec    \right)  \diag(\pvec)^{-1}  \wvec } {( \wvec' \diag(\pvec)^{-1}  \wvec)^2}  \right)
   \\
   &=   \frac 1 2 \left( \theta- \theta_0 \right)^2  
   \left( \frac 1  {\sum_{i=1}^N \sum_{t=1}^T W_{it}^2 P_{it}^{-1}} 
    +  \frac{\sum_{i=1}^N \sum_{t=1}^T W_{it}^2 P_{it}^{-2} (P_{it} - D_{it} )} {(\sum_{i=1}^N \sum_{t=1}^T W_{it}^2 P_{it}^{-1})^2} \right) 
  \\
  &= \frac{NT} 2\, c_1 \, (\nu - \nu_0)^2   ,
\end{align*}
with $c_1$ as defined in the statement of the proposition, and  $\nu - \nu_0 = (NT)^{-1}  \left( \theta- \theta_0 \right)$.

 Thus, if $M_{it}(\theta) =  D_{it}  \,   E_{it}  -  ( \nu - \nu_0)  V_{it} $
 satisfies $\| \Mvec(\theta) \|_\infty \leq \rho$, then we have
\begin{align*}
     L_{NT}(  \theta,\rho)
     & \geq
        \min_{ \left\{ \Gammavec \in \mathbb{R}^{N \times T} \, : \, \theta = \wvec' \gammavec \right\} }   Q^{({\rm low},1)}_{NT}(  \Gammavec)
          + Q^{({\rm low},2)}_{NT} 
   \\
     &=     \frac{NT} 2\, c_1 \, (\nu - \nu_0)^2
   +    \sum_{i=1}^N \sum_{t=1}^T    M_{it}(\theta) \,  \Gamma^{\infty}_{it}  
  +    \frac 1 2 \sum_{i=1}^N \sum_{t=1}^T   D_{it} \, E_{it}^2 ,
\end{align*}
and combing this with \eqref{LowerBoundLtrue} gives
\begin{align*}
  &   \frac{ L_{NT}(  \theta,\rho) -  L_{NT}(  \theta_0,\rho) } {NT}
     \geq     \frac{c_1} 2 \, (\nu - \nu_0)^2
   +   \frac 1 {NT}  \sum_{i=1}^N \sum_{t=1}^T    M_{it}(\theta) \,  \Gamma^{\infty}_{it}  
   -    \frac{ \rho }{NT} \|  \Gammavec^{\infty} \|_1 
 \\  
 & \qquad   =  \frac{c_1} 2 \, (\nu - \nu_0)^2
   +   \frac 1 {NT}  \sum_{i=1}^N \sum_{t=1}^T   D_{it} \,  E_{it} \,  \Gamma^{\infty}_{it}
    \\
    & \qquad   - (\nu - \nu_0)   \frac 1 {NT}  \sum_{i=1}^N \sum_{t=1}^T    V_{it} \,  \Gamma^{\infty}_{it}  
   -    \frac{ \rho }{NT} \|  \Gammavec^{\infty} \|_1 .
\end{align*}
Using the assumption $c_1>0$ and definitions of $c_2$ and $c_3$ in the proposition this inequality can equivalently be written as
\begin{align}
     \frac{2 \left[ L_{NT}(  NT \nu,\rho) -  L_{NT}(  NT  \nu_0, \rho) \right]} {c_1 \, NT}
  &   \geq    (\nu - \nu_0)^2
    -     \frac{2 \, c_2} {c_1} (\nu - \nu_0)
    +  \left(  \frac{c_2} { c_1}  \right)^2
   -    c_3 
  \nonumber \\
   &=  \left(\nu - \nu_0 - \frac{c_2} {c_1} \right)^2   -    c_3 .
   \label{FinalInquality}
\end{align}
Notice that $c_3>0$ because our assumptions guarantee that   $ \| \Evec \|_\infty < \rho$
and therefore $ \rho \|  \Gammavec^{\infty} \|_1 \geq  \sum_{i=1}^N \sum_{t=1}^T    E_{it} \,  \Gamma^{\infty}_{it}$, according to \eqref{DefTraceNorm}.

The inequality in \eqref{FinalInquality} was derived under the assumption that $\| \Mvec(NT \nu) \|_\infty \leq \rho$.
Define $ \nu^*_+(\varepsilon) \in \mathbb{R}$ and  $ \nu^*_-(\varepsilon) \in \mathbb{R}$ by
$$
    \nu^*_\pm(\varepsilon)  := \nu_0 \pm \left( c_4 + \varepsilon \right) ,
    \qquad
    \text{for}
    \quad
    0< \varepsilon \leq  \frac{ \rho -   \| \Evec \|_\infty - c_4  \| \Vvec \|_\infty  } { \| \Vvec \|_\infty} .
$$
Our assumption  $  \| \Evec \|_\infty +c_4  \| \Vvec \|_\infty < \rho$ guarantees that such an $ \varepsilon > 0$ exists.
Using the triangle inequality we find that
\begin{align*}
    \| \Mvec(NT  \nu^*_\pm(\varepsilon) ) \|_\infty =    \|  \Evec - ( \nu^*_\pm(\varepsilon) -\nu_0) \Vvec \|_\infty  \leq \| \Evec \|_\infty +  | \nu^*_\pm(\varepsilon) - \nu_0|  \| \Vvec \|_\infty  \leq \rho,
\end{align*}
where the final inequality follows from the definition of $ \nu^*_\pm(\varepsilon) $.
The conditions for \eqref{FinalInquality} is therefore satisfies by $\nu =  \nu^*_\pm(\varepsilon) $, that is, we have
\begin{align*}
     \frac{2 \left[ L_{NT}(  NT  \nu^*_\pm(\varepsilon),\rho) -  L_{NT}(  NT  \nu_0, \rho) \right]} {c_1 \, NT}
  &   \geq    \left( \nu^*_\pm(\varepsilon) - \nu_0 - \frac{c_2} {c_1} \right)^2   -    c_3 
  \\
  &= \left(  c_4 + \varepsilon \mp \frac{c_2} {c_1} \right)^2   -    c_3 
  \\
  &= \left(   \sqrt{c_3}   + \varepsilon  +  \frac{|c_2| \mp c_2}{c_1}  \right)^2   -    c_3 
  \\
  &\geq \left(   \sqrt{c_3}   + \varepsilon    \right)^2   -    c_3 
  \\
  &>0.
\end{align*}
where we used the definition $c_4 =  \sqrt{c_3}   +  \frac{|c_2|}{c_1} $.

$L_{NT}(  NT \nu,\rho)$ is a convex function of $\nu= \theta/NT$, because it was obtained via profiling of the convex function $Q_{NT}(  \Gammavec,\rho) $.
The  value $\nu_0$
lies in the interval  $[\nu^*_+(\varepsilon), \nu^*_-(\varepsilon)]$,
and we have shown that
$L_{NT}(  NT  \nu_0, \rho) < L_{NT}(  NT  \nu^*_\pm(\varepsilon),\rho)$.
It must therefore be the case that the optimal $\widehat \nu = NT \widehat \theta$ 
that minimizes  $L_{NT}(  NT \nu,\rho)$ also lies in the interval $[\nu^*_+(\varepsilon), \nu^*_-(\varepsilon)]$
--- otherwise we obtain a contradiction to the convexity of $L_{NT}(  NT \nu,\rho)$.
Thus, we have shown that
   \begin{align*}
         \left| \widehat \nu  - \nu_0    \right|  \leq c_4 + \varepsilon ,
    \end{align*}
and because we can choose $ \varepsilon >0$ arbitrarily small it must be the case that
   \begin{align*}
         \left| \widehat \nu  - \nu_0    \right|  \leq c_4 ,
    \end{align*}
which is what we wanted to show.

\hfill$\square$\\

\end{document}